\documentclass{JHEP3}
\usepackage[english]{babel}
\usepackage{cite}

\usepackage{epsf}
\usepackage{amssymb}
\usepackage{amsmath}
\usepackage{amsfonts}
\usepackage{psfrag,epsfig,graphicx,graphics}

\providecommand{\U}[1]{\protect\rule{.1in}{.1in}}

\def\slashchar#1{\setbox0=\hbox{$#1$}
   \dimen0=\wd0
   \setbox1=\hbox{/} \dimen1=\wd1
   \ifdim\dimen0>\dimen1
      \rlap{\hbox to \dimen0{\hfil/\hfil}}
      #1
   \else
      \rlap{\hbox to \dimen1{\hfil$#1$\hfil}}
      /
   \fi}

\def\tr{{\rm tr}}

\def\bei{\begin{itemize}}
\def\ei{\end{itemize}}

\def\beeq{\begin{eqnarray}} 
\def\beqa{\begin{eqnarray}}
\def\bea{\begin{eqnarray}}

\def\eea{\end{eqnarray}}
\def\eqa{\end{eqnarray}}
\def\eeeq{\end{eqnarray}}

\def\eqar{\end{array}}
\def\beqar{\begin{array}}

\def\beas{\begin{eqnarray*}}
\def\beqas{\begin{eqnarray*}}

\def\eqas{\end{eqnarray*}}
\def\eeas{\end{eqnarray*}}

\def\beq{\begin{equation}} 
\def\be{\begin{equation}}

\def\ee{\end{equation}}
\def\eq{\end{equation}}
\def\eeq{\end{equation}}

\def\beqd{\begin{displaymath}}
\def\eeqd{\end{displaymath}}
\def\eqd{\end{displaymath}}

\def\beeq{\begin{eqnarray}} \def\eeeq{\end{eqnarray}}

\newcommand{\fin}{\end{document}}

\title{Impact factor for high-energy two and three jets diffractive production}
\author{R.~Boussarie\\
LPT, Universit{\'e} Paris-Sud, CNRS, 91405, Orsay, France\\
Email: \email{Renaud.Boussarie@th.u-psud.fr}}
\author{A.~V.~Grabovsky\\
Budker Institute of Nuclear Physics and Novosibirsk State University, 630090 Novosibirsk, Russia\\
Email: \email{A.V.Grabovsky@inp.nsk.su}}
\author{ L. Szymanowski\\
National Centre for Nuclear Research (NCBJ), Warsaw, Poland\\
Email: \email{Lech.Szymanowski@fuw.edu.pl}}
\author{S. Wallon\\
LPT, Universit{\'e} Paris-Sud, CNRS, 91405, Orsay, France {\em \&} \\
UPMC Univ. Paris 06, facult\'e de physique, 4 place Jussieu, 75252 Paris Cedex 05, France\\
Email: \email{Samuel.Wallon@th.u-psud.fr}}

\abstract{We present the calculation of the impact factor for  the $\gamma^{(*)}\to q\bar{q}g$ transition within  Balitsky's high energy operator expansion. We also rederive  the impact factor for  the $\gamma^{(*)}\to q\bar{q}$ transition within the same framework. These results provide the necessary building blocks for further phenomenological studies of inclusive diffractive deep inelastic scattering as well as for two and three jets diffractive production which go beyond approximations discussed in the litterature.}

\begin{document}

\pagestyle{empty}
\newpage

\mbox{}

\pagestyle{plain}

\setcounter{page}{1}
\section{Introduction}
\label{Sec:Intro}

Diffraction is one of the key tools to understand the dynamics of strong interaction, and it has been studied since the sixties. In particular, the research program performed at HERA has shown that semi-hard diffractive processes, in which a hard scale allows one to deal with 
QCD in its perturbative regime, could provide a quantitative lever-arm to understand the internal dynamics of the nucleon 
in a regime of very high gluon densities\footnote{For reviews, see refs.~\cite{Wusthoff:1999cr,Wolf:2009jm}}.
Among the whole set of $\gamma^* p \to X$ deep inelastic scattering (DIS) events, almost 10 \%  reveal a rapidity gap between the proton remnants 
and the hadrons 
coming from the fragmentation region of the initial virtual photon, so that the process looks rather like 
$\gamma^* p \to X  \, Y$~\cite{Aktas:2006hx,Aktas:2006hy,Chekanov:2004hy,Chekanov:2005vv,Aaron:2010aa,Aaron:2012ad,Chekanov:2008fh,Aaron:2012hua}, where $Y$ is the outgoing proton or one of its low-mass excited states. This subset of events is called diffractive deep inelastic scattering (DDIS). DDIS can be studied at the 
inclusive level, or further analyzed by considering diffractive jet production as well as exclusive meson production. 
One of the main cornerstones of diffraction is the concept of Pomeron, which carries the quantum numbers of vacuum and 
which is exchanged at high energy between $X$ and $Y$. Diffraction can be described according to several approaches, important for phenomenological applications. In the perturbative QCD approach, justified by the existence of a hard scale (like the photon virtuality $Q^2$), one can rely on a QCD factorization theorem~\cite{Collins:1997sr}. This is the essence of the first approach, which involves
a {\em resolved} Pomeron contribution:  the diffractive structure function is expressed as the convolution of a coefficient function with a diffractive parton distribution, which is analogous to the usual parton distribution function (PDF), but with the proton replaced by a Pomeron.

Besides, at high energies, it is natural to model the diffractive events by a {\em direct} Pomeron contribution involving the coupling of a Pomeron with the diffractive state. The diffractive states can be modelled in perturbation theory by a  $q \bar{q}$ pair (for moderate $M^2$, where $M$ is the invariant mass of the  diffractively produced state $X$) or a $q \bar{q} g$ state for larger values of $M^2$. Based on such a model, with
 a two-gluon exchange picture for the Pomeron,  a good description of HERA data for diffraction could be achieved
\cite{Bartels:1998ea}. One of the important features of this approach is that the $q \bar{q}$ component with a longitudinally polarized photon plays a crucial role in the region of small
diffractive mass $M$, although it is a
twist-4 contribution.
A further analysis was then performed, combining both the resolved and the direct components~\cite{Martin:2005hd,Martin:2006td}, including a $q \bar{q}$ exchange on top of the gluon pair to model the Pomeron. 

In the direct components considered there, the $q \bar{q} g$ diffractive state has been studied in two particular limits. The first one, valid for very large $Q^2$, corresponds to a collinear approximation in which the transverse momentum of the gluon is assumed to be much smaller than the transverse momentum of the emitter~\cite{Wusthoff:1995hd,Wusthoff:1997fz}. This approximation allows one to extract the leading logarithm in $Q^2$, based on the strong ordering of transverse momenta typical of DGLAP evolution~\cite{Gribov:1972ri, Lipatov:1974qm, Altarelli:1977zs, Dokshitzer:1977sg}. The second one~\cite{Bartels:1999tn,Bartels:2002ri}, valid for very large $M^2$, is based on the assumption of a strong ordering of longitudinal momenta, encountered in BFKL equation~\cite{Fadin:1975cb, Kuraev:1976ge, Kuraev:1977fs, Balitsky:1978ic}.

The main aim of the present article is to compute the $\gamma^* \to q \bar{q} g$ impact factor and to rederive the $\gamma^* \to q \bar{q}$ impact factor, both at tree level, with an arbitrary
number of $t-$channel gluons, here described within the Wilson line formalism, also called QCD shockwave approach~\cite{Balitsky:1995ub, Balitsky:1998kc, Balitsky:1998ya, Balitsky:2001re}. In particular, the 
$\gamma^* \to q \bar{q} g$ transition is computed without any soft or collinear approximation for the emitted gluon, in contrast with the above mentioned calculations. These results provide necessary generalization of buiding blocks for inclusive DDIS as well as for two- and three-jet diffractive production. Since the results we derived can account for an arbitrary number of $t-$channel gluons, this could allow to include higher twist effects which are suspected to be rather important in DDIS for $Q^2 \lesssim 5$ GeV$^2$~\cite{Motyka:2012ty}.

The QCD shock-wave approach on which we rely is an operator language based on the concept of
factorization of the scattering amplitude in rapidity space and on the extension
to high-energy (Regge limit) of the Operator Product Expansion (OPE) technique, 
which was only known at moderate energy (Bjorken limit) before, as an expansion in terms of local operators or in terms of light-ray operators~\cite{Balitsky:1987bk}. In DIS off a hadron at high-energy, the matrix elements made of Wilson line operators appearing in the OPE
describe the non perturbative part of the process, and their evolution in rapidity 
is related to the evolution of the structure function of the target. The evolution equation can be obtained relying on background field techniques. 
The Wilson-line operators in the high-energy OPE evolve with respect to rapidity according to the Balitsky equation, which reduces to the Balitsky-Kovchegov (BK) 
equation~\cite{Balitsky:1995ub, Balitsky:1998kc, Balitsky:1998ya, Balitsky:2001re, Kovchegov:1999yj, Kovchegov:1999ua} in the large $N_c$ limit.
According to the best of our knowledge, this shock-wave approach was only used for evolution equations and for impact factors at inclusive level, namely only for the $\gamma^* \to \gamma^*$ impact factor at next-to-leading order~\cite{Balitsky:2010ze,Balitsky:2012bs}.
Its application shows that this method is very powerful \cite{Chirilli:2010mw} when compared with usual methods based on summation of contributions of individual Feynman diagrams computed in momentum space.

When describing a diffractive process, the above mentioned approach is natural in order to implement saturation
effects at high energies, since it is formulated in the coordinate space.
Indeed, in the dipole picture~\cite{Mueller:1989st,Nikolaev:1990ja}, when probing a nucleon with a virtual photon in the rest frame of the nucleon, due to the 
long life-time of the virtual $q \bar{q}$ pair produced by the $\gamma^*$ probe with respect to its scattering time, this pair is almost frozen during its interaction. The inclusive cross-section (for DIS) as well as the scattering amplitude (for DDIS)  thus naturally factorizes in the coordinate space in terms of an impact factor involving a dipole of given transverse size $r$ convoluted with an effective dipole-nucleon cross-section $\sigma(x,r)$, a function of Bjorken $x$ and $r$. The same picture was extended 
to the $q \bar{q} g$ intermediate state, at least in the collinear approximation in which case this intermediate state can be considered
as a gluon-gluon dipole~\cite{Wusthoff:1995hd,Wusthoff:1997fz}, the $q \bar{q}$ being  an effective gluon due to its localization in the transverse coordinate space (since the relative transverse momentum of this pair is large with respect to the transverse momentum of the emitted gluon). A step further in this spirit
was done in the case of vector meson electroproduction at twist 3, including the genuine twist 3 contribution which involves a $q \bar{q} g$ intermediate state~\cite{Anikin:2009bf,Anikin:2009hk}, for which a dipole picture was also obtained~\cite{Besse:2012ia},  based on QCD equations of motion.

This dipole picture provides the natural framework for the formulation of saturation. Indeed, the transverse size $r$ of the dipole is the natural parameter in order to implement both color transparency (for small $r$) and saturation (for large $r$). The analysis of low-$x$ saturation dynamics of the nucleon target was first introduced in refs.~\cite{GolecBiernat:1998js,GolecBiernat:1999qd} by Golec-Biernat and W\"usthoff (GBW) to describe the inclusive and diffractive structure functions of DIS. This is an additionnal reason to rely on the shock-wave analysis, which naturaly provides a tool to evaluate the $\gamma^* \to q \bar{q}$ and 
$\gamma^* \to q \bar{q} g$ impact factors in transverse coordinate space.

The paper is organized as follows. The section~\ref{Sec:Def} contains the definitions and necessary intermediate results. Section~\ref{Sec:qqbar} briefly reproduces the leading order (LO) $\gamma^* \rightarrow q\bar{q}$ impact factor. In section~\ref{Sec:qqbarg} we give the general expression for the
$\gamma\rightarrow q\bar{q}g$ impact factor, which is then calculated in sections~\ref{Sec:qqbarg-through-sw} and \ref{Sec:qqbarg-no-sw}. Section~\ref{Sec:dipole} discusses the linearized impact
factor for interaction with the color dipole. Section~\ref{Sec:momentum} is devoted to the
impact factor in the momentum space. Section~\ref{Sec:Conclusion} summarizes obtained results. Two
appendices comprise necessary technical details.

\section{Definitions and necessary intermediate results}
\label{Sec:Def}

Throughout this paper, we use the following notations. We introduce the light cone vectors $n_{1}$ and
$n_{2}$%
\begin{equation}
n_{1}=\left(  1,0,0,1\right)  ,\quad n_{2}=\frac{1}{2}\left(  1,0,0,-1\right)
,\quad n_{1}^{+}=n_{2}^{-}=n_{1} \cdot n_{2}=1 \,,
\end{equation}
and for any vector $p$ we have%
\begin{equation}
p^{+}=p_{-}=p \cdot n_{2}=\frac{1}{2}\left(  p^{0}+p^{3}\right)  ,\qquad p_{+}%
=p^{-}=p \cdot n_{1}=p^{0}-p^{3},
\end{equation}%
\begin{equation}
p=p^{+}n_{1}+p^{-}n_{2}+p_{\bot},\qquad p^{2}=2p^{+}p^{-}-\vec{p}^{\,2},
\end{equation}%
\begin{equation}
\quad p\cdot k=p^{\mu}k_{\mu}=p^{+}k^{-}+p^{-}k^{+}-\vec{p}\cdot \vec{k}=p_{+}%
k_{-}+p_{-}k_{+}-\vec{p}\cdot \vec{k}.
\end{equation}
The derivatives and the metric tensor have the form%
\begin{equation}
\partial_{\pm}=\partial^{\mp}=\frac{\partial}{\partial z^{\pm}}=\frac
{\partial}{\partial z_{\mp}},\qquad\partial_{i}=-\partial^{i}=\frac{\partial
}{\partial z^{i}}=-\frac{\partial}{\partial z_{i}},
\end{equation}%
\begin{equation}
g^{\mu\nu}=g_{\mu\nu}=%
\begin{pmatrix}
0 \ \ & 1 & \mbox{ }\mbox{ }0 & \mbox{ }\mbox{ }0\\
1 \ \ & 0 & \mbox{ }\mbox{ }0 & \mbox{ }\mbox{ }0\\
0 \ \ & 0 & -1 & \mbox{ }\mbox{ }0\\
0 \ \ & 0 & \mbox{ }\mbox{ }0 & -1
\end{pmatrix}
,\qquad\text{the indices\thinspace are }+,-\,,1\,,2;
\end{equation}%
\begin{equation}
\epsilon^{+\beta\gamma-}=-\epsilon^{0\beta\gamma3}=-e^{\beta\gamma3}.
\end{equation}
We denote the initial photon momentum as $k,$ and the outgoing quark, antiquark,
and gluon momenta as $p_{q}$, $p_{\bar{q}},$ and $p_{g}.$ The corresponding
longitudinal momentum fractions are%
\begin{equation}
\frac{p_{q}^{+}}{k^{+}}=x_{q}\,,\quad\frac{p_{\bar{q}}^{+}}{k^{+}}=x_{\bar{q}%
}\,,\quad\frac{p_{g}^{+}}{k^{+}}=x_{g}\,.\label{xs}%
\end{equation}
For simplicity, we work with a photon in the forward kinematics%
\begin{equation}
\vec{k}=0,\quad k^{\mu}=k^{+}n_{1}^{\mu}+\frac{k^{2}}{2k^{+}}n_{2}^{\mu}%
,\quad-k^{2}=Q^{2}>0.\label{photonk}%
\end{equation}
Its longitudinal and transverse polarization vectors read%
\begin{equation}
\varepsilon_{L}^{\alpha}=\frac{1}{\sqrt{-k^{2}}}\left(  k^{+}n_{1}^{\alpha
}-\frac{k^{2}}{2k^{+}}n_{2}^{\alpha}\right)  ,\quad\varepsilon_{L}^{+}%
=\frac{k^{+}}{Q},\quad\varepsilon_{L}^{-}=\frac{Q}{2k^{+}},\label{eL}%
\end{equation}%
\begin{equation}
\varepsilon_{T}^{\alpha}=\varepsilon_{T\bot}^{\alpha}=\frac{1}{\sqrt{2}%
}(0,1,i\,s,0),\quad s=\pm1.\label{eT}%
\end{equation}
Here $s$ is the helicity of the photon. For the outgoing gluon we work in the
light cone gauge $A\cdot n_{2}=0.$ Therefore
\begin{equation}
\varepsilon_{g\nu}^{\ast}=\frac{(\vec{\varepsilon}_{g}^{\,\,\ast}\cdot \vec{p}_{g}%
)}{p_{g}^{+}}n_{2\nu}+\varepsilon_{g\bot\nu}^{\ast}=\left(  g_{\bot\nu\alpha
}-\frac{p_{g\bot\alpha}n_{2\nu}}{p_{g}^{+}}\right)  \varepsilon_{g}%
^{\ast\alpha},\label{guageProector}%
\end{equation}
and we can use the same transverse polarization vectors as for the photon
\begin{equation}
\varepsilon_{g\bot}^{\ast\alpha}=\frac{1}{\sqrt{2}}(0,1,-i \, s_{g},0),\quad
s_{g}=\pm1.\label{eg}%
\end{equation}
It is convenient to introduce the following vectors%
\begin{equation}
\vec{P}_{\bar{q}}{}=\frac{\vec{p}_{g}{}}{x_{g}}-\frac{\vec{p}_{\bar{q}}{}%
}{x_{\bar{q}}},\quad\vec{P}_{q}=\frac{\vec{p}_{g}{}}{x_{g}}-\frac{\vec{p}_{q}%
}{x_{q}}.\label{PP}%
\end{equation}
Then%
\begin{equation}
(\vec{P}_{\bar{q}}{}\cdot\vec{\varepsilon}_{g}^{\,\,\ast})=\frac{p_{\bar{q}%
}\cdot\varepsilon_{g}^{\ast}}{x_{\bar{q}}},\quad(\vec{P}_{q}{}\cdot \vec{\varepsilon}%
_{g}^{\,\,\ast})=\frac{p_{q}\cdot\varepsilon_{g}^{\ast}}{x_{q}}.
\end{equation}
To simplify the vector products with the polarization vectors we will use the
following identities%
\begin{equation}
\lbrack\vec{a}\times\vec{\varepsilon}_{T}]=i \, s(\vec{a}\cdot \vec{\varepsilon}%
_{T}),\quad\lbrack\vec{a}\times\vec{\varepsilon}_{g}^{\,\,\ast}]=-i \,s_{g}%
(\vec{a}\cdot \vec{\varepsilon}_{g}^{\,\,\ast})\,,\label{crossToSP}%
\end{equation}
where
$[\vec
{a}\times\vec{b}] \equiv e^{\gamma\beta3}a^{\gamma}b^{\beta}\,.$
The fermion propagator in the shock wave background can be read from ref.~\cite{Balitsky:1995ub} and is given by
\[
G(z_{1},z_{2})=\theta(z_{1}^{+}z_{2}^{+})G_{0}\left(  z_{12}\right)  -\int
d^{4}z_{3}\delta(z_{3}^{+})G_{0}\left(  z_{13}\right)  \gamma^{+}%
\,G_{0}\left(  z_{32}\right)
\]%
\begin{equation}
\times\left(  \theta(z_{1}^{+})\theta(-z_{2}^{+})U_{\vec{z}_{3}}+\theta
(-z_{1}^{+})\theta(z_{2}^{+})U_{\vec{z}_{3}}^{\dag}\right)
\,.\label{qpropinSW}%
\end{equation}
Here $z_{ij}=z_{i}-z_{j}\,.$ The free quark propagator reads%
\begin{equation}
G_{0}(x)=\frac{2i}{\left(  2\pi\right)  ^{2}}\frac{\hat{z}}{\left(
z^{2}-i0\right)  ^{2}},\quad G_{0}\left(  p\right)  =\frac{i\hat{p}}{p^{2}%
+i0},
\end{equation}
and the Wilson lines%
\begin{equation}
U_{i}=U_{\vec{z}_{i}}=U\left(  \vec{z}_{i},\eta\right)  =P \exp\left[{ig\int_{-\infty
}^{+\infty}b_{\eta}^{-}(z_{i}^{+},\vec{z}_{i}) \, dz_{i}^{+}}\right]\label{WL}%
\end{equation}
are integrated along the path $z^{-}=0$. The operator $b_{\eta}^{-}$ is the external
shock-wave field built from only slow gluons 
which momenta are limited by the longitudinal cut-off defined by the rapidity $\eta$
\begin{equation}
b_{\eta}^{-}=\int\frac{d^{4}p}{\left(  2\pi\right)  ^{4}}e^{-ip \cdot z}b^{-}\left(
p\right)  \theta(e^{\eta}-|p^{+}|).\label{cutoff}%
\end{equation}
We use the light cone gauge
\begin{equation}
\mathcal{A}\cdot n_{2}=0,\label{gauge}%
\end{equation}
with $\mathcal{A}$ being the sum of the external field $b$ and the quantum field
$A$%
\begin{equation}
\mathcal{A=}A+b,\quad b^{\mu}\left(  z\right)  =b^{-}(z^{+},\vec{z}\,) \,n_{2}%
^{\mu}=\delta(z^{+})B\left(  \vec{z}\,\right)  n_{2}^{\mu}\,.\label{b}%
\end{equation}
Using the LSZ reduction formulas for the propagators from ref.~\cite{Balitsky:1995ub} or
summing the diagrams in this external shockwave field as in ref.~\cite{Balitsky:1995ub}
one can get the external fermion lines in the shockwave background%
\begin{equation}
\label{uinSW}
\bar{u}(p,y)|_{0>y^{+}}=\int d^{4}z\,\delta(z^{+})e^{ip \cdot z}\frac{\bar
{u}_{p}}{\sqrt{2p^{+}}}\gamma^{+}U_{\vec{z}}\,G_{0}\left(  z-y\right)
,%
\end{equation}%
\begin{equation}
\label{vinSW}
v(p,y)|_{0>y^{+}}=-\int d^{4}z\,\delta(z^{+})e^{ip \cdot z}G_{0}\left(  y-z\right)
U_{\vec{z}}^{\dag}\,\gamma^{+}\frac{v_{p}}{\sqrt{2p^{+}}}\,.
\end{equation}
In the same way one can get the gluon external line in the shockwave
background%
\begin{equation}
\epsilon_{\nu}^{\ast}(p,y)|_{0>y^{+}}=-4p^{+}\varepsilon^{\ast\alpha}%
\theta(p^{+})\int\frac{d^{4}z\,\delta\left(  z^{+}\right)  }{(2\pi)^{2}%
}\,\,\frac{e^{ip \cdot z}}{\sqrt{2p^{+}}}\,\,\,U_{\vec{z}}\frac{1}{\frac
{\partial\,\,\,\,}{\partial y^{-}}}\,\,\,\frac{g_{\bot\alpha\nu}%
(-y^{+})-(z-y)_{\bot\alpha}n_{2\nu}}{\left(  (z-y)^{2}-i0\right)  ^{2}%
}\,,\label{einSW}%
\end{equation}
where we have introduced the notation
\begin{equation}
\frac{1}{\frac{\partial\,\,\,\,}{\partial y}}f\left(  y\right)  =\int\frac
{dp}{2\pi}\int du\frac{e^{-ip\left(  y-u\right)  }}{-ip}f\left(  u\right)
,\label{1/d}%
\end{equation}
In eqs.~(\ref{uinSW}, \ref{vinSW}), the Wilson line is in the adjoint representation. If $U\rightarrow1,$ then%
\begin{equation}
\bar{u}(p,y)|_{0>y^{+}}\rightarrow\theta(p^{+})\frac{\bar{u}_{p}%
}{\sqrt{2p^{+}}}e^{ip \cdot y},\quad v(p,y)|_{0>y^{+}}\rightarrow\theta(p^{+}%
)\frac{v_{p}}{\sqrt{2p^{+}}}e^{ip \cdot y},
\end{equation}%
\begin{equation}
\epsilon_{\nu}^{\ast}(p,y)|_{0>y^{+}}\rightarrow\frac{\theta(p^{+}%
)\varepsilon_{p}^{\ast\alpha}}{\sqrt{2p^{+}}}\left(  g_{\bot\alpha\nu}%
-\frac{p_{\bot\alpha}n_{2\nu}}{p^{+}}\right)  e^{ip \cdot y}=\theta(p^{+}%
)\frac{\varepsilon_{p\nu}^{\ast}}{\sqrt{2p^{+}}}e^{ip \cdot y},\label{efree}%
\end{equation}
and we recover the results without the shockwave.

Below, we will need the following integral\footnote{The factor $-i$ 
in eq.~(\ref{vertexIntegral}) corrects a misprint from ref.~\cite{Balitsky:2012bs}.} derived in ref.~\cite{Balitsky:2012bs}%
\beqa
&&\hspace{-.7cm}\int d^{4}z\frac{\hat{x}-\hat{z}}{(x-z)^{4}}\left[  \gamma^{\mu}\frac{z^{\nu}%
}{z^{4}}-\gamma^{\nu}\frac{z^{\mu}}{z^{4}}\right]  \frac{\hat{z}-\hat{y}%
}{(z-y)^{4}}=\frac{-i\pi^{2}}{x^{2}y^{2}(x-y)^{2}}\left(  \frac{\hat{x}%
\gamma^{\nu}\hat{y}x^{\mu}-\hat{x}\gamma^{\mu}\hat{y}x^{\nu}}{x^{2}}\right.
\nonumber
\\
&&\hspace{-1.4cm}\left.  +\frac{\hat{x}\gamma^{\nu}\hat{y}y^{\mu}-\hat{x}\gamma^{\mu}\hat
{y}y^{\nu}}{y^{2}}+\frac{(\gamma^{\nu}\gamma^{\mu}-\gamma^{\mu}\gamma^{\nu
})\hat{y}}{2}+\frac{\hat{x}\left(  \gamma^{\mu}\gamma^{\nu}-\gamma^{\nu}%
\gamma^{\mu}\right)  }{2}+2\frac{y^{\mu}x^{\nu}-y^{\nu}x^{\mu}}{(x-y)^{2}%
}\left[  \hat{y}-\hat{x}\right]  \right) \! , 
\label{vertexIntegral}
\eqa
and the following representation of the McDonald functions%
\begin{equation}
-2K_{0}\left(  2\sqrt{ab}\right)  =\int_{-\infty}^{0}\frac{dz}{z}%
e^{i(a-i0)z-i\frac{b+i0}{z}},\quad K_{1}(r)=-K_{0}^{\prime}(r)\,, \label{Kint}%
\end{equation}
together with the fact that they obey the Bessel equation%
\begin{equation}
\Delta K_{0}(r)=\left(  \frac{\partial^{2}}{\partial r^{2}}+\frac{1}{r}%
\frac{\partial}{\partial r}\right)  K_{0}(r)=K_{0}(r)+2\pi\delta\left(
\vec{r}\right)  . \label{Klaplasian}%
\end{equation}
We will also need the following Dirac structures%
\begin{equation}
\bar{u}_{p_{q}}\gamma^{+}\,\,v_{p_{_{\bar{q}}}}=\lambda_{q}\bar
{u}_{p_{q}}\gamma^{+}\,\gamma^{5}\,v_{p_{_{\bar{q}}}}=\delta_{\lambda
_{q},-\lambda_{\bar{q}}}\sqrt{2p_{q}^{+}2p_{\bar{q}}^{+}},\quad\lambda_{q}%
=\pm1, \label{ug+v}%
\end{equation}%
\begin{equation}
\bar{u}_{p_{q}}\gamma^{j}v_{p_{_{\bar{q}}}}=\lambda_{q}\overline
{u}_{p_{q}}\gamma^{j}\gamma^{5}v_{p_{_{\bar{q}}}}=\delta_{\lambda_{q}%
,-\lambda_{\bar{q}}}\sqrt{2p_{\bar{q}}^{+}2p_{q}^{+}}\left(  \frac{p_{q}^{j}%
}{2p_{q}^{+}}+\frac{p_{\bar{q}}^{j}}{2p_{\bar{q}}^{+}}+i\varepsilon
^{tj3}\lambda_{q}\left(  \frac{p_{q}^{t}}{2p_{q}^{+}}-\frac{p_{\bar{q}}^{t}%
}{2p_{\bar{q}}^{+}}\right)  \right)  , \label{ugjv}%
\end{equation}
where $\lambda_{q}$ is the quark helicity.

\section{LO impact factor for $\gamma\rightarrow q\bar{q}$ transition}%
\label{Sec:qqbar}

\begin{figure}[th]%
\centering
\includegraphics[
height=6.5781cm,
width=10.8451cm
]%
{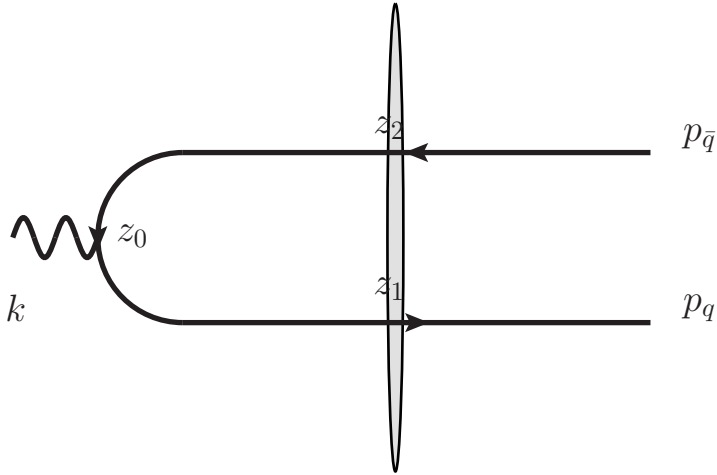}%
\caption{LO $\gamma\rightarrow q\bar{q}$ impact factor}%
\label{lo}%
\end{figure}
In this section  we will briefly reproduce the known expression for the
LO $\gamma^*\rightarrow q\bar{q}$\ impact factor, see fig.~\ref{lo}, for completeness of this paper. The LO matrix element for the
electromagnetic current in the shockwave background reads%
\begin{equation}
\tilde{M}_{0}^{\alpha}=\delta_{l}^{n}\frac{\langle0|b_{p_{\bar{q}}}%
^{l}(a_{p_{q}})_{n}\overline{\psi}\left(  z_{0}\right)  \gamma^{\alpha}%
\psi\left(  z_{0}\right)  e^{i\int\mathcal{L}_{i}\left(  z\right)
dz}|0\rangle}{\langle0|e^{i\int\mathcal{L}\left(  z\right)  dz}|0\rangle}=\int
d\vec{z}_{1}d\vec{z}_{2}F\left(  p_{q},p_{\bar{q}},z_{0},\vec{z}_{1},\vec
{z}_{2}\right)  ^{\alpha}\tr(U_{1}U_{2}^{\dag}).
\end{equation}
Here $a$ and $b$ are the quark and antiquark annihilation operators carrying colour indices $n$ and $l$,%
\beqa
\label{defF}
&&F\left(  p_{q},p_{\bar{q}},z_{0},\vec{z}_{1},\vec{z}_{2}\right)  ^{\alpha
}=\int dz_{1}^{-}dz_{1}^{+}\delta(z_{1}^{+})\int dz_{2}^{-}dz_{2}^{+}%
\delta(z_{2}^{+})\frac{e^{ip_{q} \cdot z_{1}+ip_{_{\bar{q}}} \cdot z_{2}}}{\sqrt{2p_{q}%
^{+}2p_{_{\bar{q}}}^{+}}}%
\nonumber
\\
&&\times\bar{u}_{p_{q}}\gamma^{+}\,G_{0}\left(  z_{10}\right)
\gamma^{\alpha}G_{0}\left(  z_{02}\right)  \gamma^{+}\,v_{p_{_{\bar{q}}}%
}=\frac{\theta(p_{q}^{+})\theta(p_{\bar{q}}^{+})e^{-i\vec{p}_{q} \cdot \vec{z}%
_{1}-i\vec{p}_{_{\bar{q}}} \cdot \vec{z}_{2}}}{4(2\pi)^{4}(z_{0}^{+})^{2}\sqrt
{2p_{q}^{+}2p_{_{\bar{q}}}^{+}}}%
\nonumber
\\
&&\times\bar{u}_{p_{q}}\gamma^{+}\,\left(ip_{q}^{+}\gamma^{-}-\gamma_{\bot
}^{\beta}\frac{\partial}{\partial z_{1\bot}^{\beta}}\right)\gamma^{\alpha}%
\left(ip_{\bar{q}}^{+}\gamma^{-}-\gamma_{\bot}^{\gamma}\frac{\partial}{\partial
z_{2\bot}^{\gamma}}\right)\gamma^{+}\,v_{p_{_{\bar{q}}}}
\nonumber
\\
&&\exp\left[ip_{q}^{+}\left(z_{0}%
^{-}+\frac{-z_{10\bot}^{\,\,\,2}+i0}{-2z_{0}^{+}}\right)\right] 
\exp\left[ip_{_{\bar{q}}}%
^{+}\left(z_{0}^{-}+\frac{-z_{20\bot}^{\,\,\,2}+i0}{-2z_{0}^{+}}\right)\right].
\eqa
We now introduce $M_{0}^{\alpha}\,,$ built from $\tilde{M}_{0}^{\alpha}$
by substracting the non-interacting term, i.e.
\begin{equation}
M_{0}^{\alpha}=\int d\vec{z}_{1}d\vec{z}_{2}F\left(  p_{q},p_{\bar{q}}%
,z_{0},\vec{z}_{1},\vec{z}_{2}\right)  ^{\alpha}(\tr(U_{1}U_{2}^{\dag})-N_{c})\,.
\label{M0int}%
\end{equation}
The Fourier transform of $F$ w.r.t. $z_{0}$ is defined as
\begin{equation}
F\left(  p_{q},p_{\bar{q}},k,\vec{z}_{1},\vec{z}_{2}\right)  ^{\alpha}=\int
d^{4}z_{0} \, e^{-ik \cdot z_{0}}F\left(  p_{q},p_{\bar{q}},z_{0},\vec{z}_{1},\vec{z}%
_{2}\right)  ^{\alpha}.
\end{equation}
In  the kinematics  (\ref{photonk}) we chose for the photon, we have%
\beqa
\label{Fphoton}
&&F\left(  p_{q},p_{\bar{q}},k,\vec{z}_{1},\vec{z}_{2}\right)  ^{\alpha}%
=\theta(p_{q}^{+})\theta(p_{\bar{q}}^{+})\frac{i\delta\left(  k^{+}-p_{q}%
^{+}-p_{\bar{q}}^{+}\right)  }{2k^{+}(2\pi)^{2}\sqrt{2p_{q}^{+}2p_{_{\bar{q}}%
}^{+}}}e^{-i\vec{p}_{q} \cdot \vec{z}_{1}-i\vec{p}_{_{\bar{q}}} \cdot \vec{z}_{2}}%
\nonumber
\\
&&\times\bar{u}_{p_{q}}\gamma^{+}\,\left(ip_{q}^{+}\gamma^{-}-\gamma_{\bot
}^{\beta}\frac{\partial}{\partial z_{1\bot}^{\beta}}\right)\gamma^{\alpha}%
\left(ip_{\bar{q}}^{+}\gamma^{-}-\gamma_{\bot}^{\gamma}\frac{\partial}{\partial
z_{2\bot}^{\gamma}}\right)\gamma^{+}\,v_{p_{_{\bar{q}}}}K_{0}\left(Q\sqrt{x_{q}x_{\bar
{q}}\vec{z}_{12}^{\,\,2}}\right) \,.
\eqa
Calculating the derivatives with $\alpha=-$\ we will encounter%
\beqa
&&g_{\bot}^{\gamma\beta}\frac{\partial}{\partial z_{2\bot}^{\gamma}}%
\frac{\partial}{\partial z_{1\bot}^{\beta}}\,K_{0}\left(Q\sqrt{x_{q}x_{\bar{q}}%
\vec{z}_{12}^{\,\,2}}\right)=\Delta_{\vec{z}_{12}}K_{0}\left(Q\sqrt{x_{q}x_{\bar{q}}%
\vec{z}_{12}^{\,\,2}}\right) \nonumber\\
&&=Q^{2}x_{q}x_{\bar{q}}K_{0}\left(Q\sqrt{x_{q}x_{\bar{q}}%
\vec{z}_{12}^{\,\,2}}\right)+2\pi\delta(\vec{z}_{12}),
\eqa
where we used (\ref{Klaplasian}) to derive this expression. However, the  term with the $\delta$ distribution will give no contribution to (\ref{M0int}), therefore we can
drop it. That being done, we get%
\beqa
\label{FL}
&&F\left(  p_{q},p_{\bar{q}},k,\vec{z}_{1},\vec{z}_{2}\right)  ^{\alpha
}\varepsilon_{L\alpha}=\theta(p_{q}^{+})\,\theta(p_{\bar{q}}^{+})\frac
{\delta\left(  k^{+}-p_{q}^{+}-p_{\bar{q}}^{+}\right)  }{(2\pi)^{2}}%
e^{-i\vec{p}_{q}\cdot \vec{z}_{1}-i\vec{p}_{_{\bar{q}}}\cdot\vec{z}_{2}}\nonumber
\\
&&\times(-2i)\delta_{\lambda_{q},-\lambda_{\bar{q}}}\,x_{q}x_{\bar{q}}%
\,Q\,K_{0}\left(Q\sqrt{x_{q}x_{\bar{q}}\vec{z}_{12}^{\,\,2}}\right)
\eqa
and
\beqa
\label{FT}
&&F\left(  p_{q},p_{\bar{q}},k,\vec{z}_{1},\vec{z}_{2}\right)  ^{j}%
\varepsilon_{Tj}=\theta(p_{q}^{+})\,\theta(p_{\bar{q}}^{+})\frac{\delta\left(
k^{+}-p_{q}^{+}-p_{\bar{q}}^{+}\right)  }{(2\pi)^{2}}e^{-i\vec{p}_{q}\cdot\vec
{z}_{1}-i\vec{p}_{_{\bar{q}}}\cdot\vec{z}_{2}}
\nonumber
\\
&&\times\delta_{\lambda_{q},-\lambda_{\bar{q}}}\left(  x_{q}-x_{\bar{q}%
}+s\lambda_{q}\right)  \frac{\vec{z}_{12} \cdot \vec{\varepsilon}_{T}}{\vec{z}_{12}^{\,\,2}}
Q\sqrt
{x_{q}x_{\bar{q}}\vec{z}_{12}^{\,\,2}} K_{1}\left(Q\sqrt{x_{q}x_{\bar{q}%
}\vec{z}_{12}^{\,\,2}}\right).
\eqa
Using the identity
\begin{equation}
\label{Fplus}
F^{+}=F^{-}\frac{2(k^{+})^{2}}{Q^{2}}\,,
\end{equation}
one can easily check that the electromagnetic gauge invariance 
\begin{equation}
F\left(  p_{q},p_{\bar{q}},k,\vec{z}_{1},\vec{z}_{2}\right)  ^{\alpha
}k_{\alpha}=0
\end{equation}
is satisfied. The results (\ref{FL}) and (\ref{FT}) are consistent with the well known result for the $\gamma\rightarrow q\bar{q}$ wave-function which was derived for example in ref.~\cite{Ivanov:1998jw}.

\section{General expression for $\gamma\rightarrow q\bar{q}g$ impact factor}%
\label{Sec:qqbarg}

\begin{figure}[th]%
\centering
\includegraphics[height=10cm
]%
{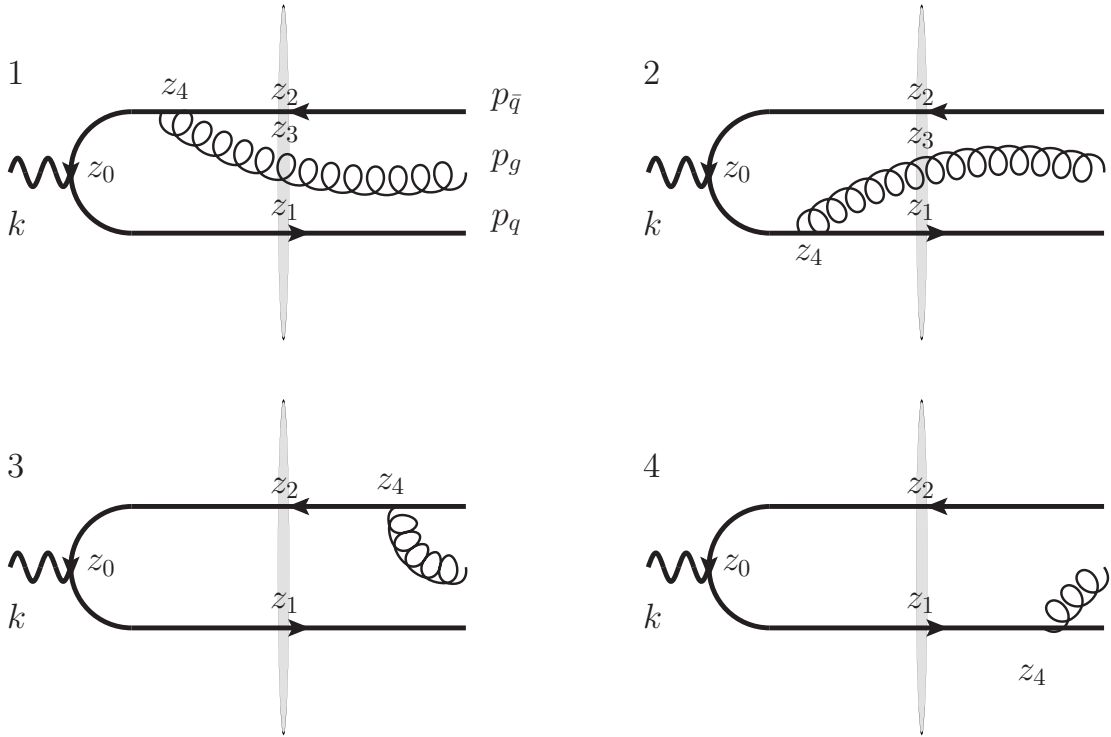}%
\caption{Impact factor for 3 jet production. The grey ellipse stands for the
shockwave at $z^{+}=0$. The lines crossing the ellipse are calculated in the
shock wave background (\ref{qpropinSW}), (\ref{uinSW}--\ref{einSW}).}%
\label{3jetif}%
\end{figure}

We will extract the impact factor from the following matrix element%
\beqa
\tilde{M}^{\alpha}&=&(t^{b})_{l}^{n}\frac{\langle0|c^b_{p_{g}}b_{p_{\bar{q}}}%
^{l}(a_{p_{q}})_{n}\overline{\psi}\left(  z_{0}\right)  \gamma^{\alpha}%
\psi\left(  z_{0}\right)  e^{i\int\mathcal{L}_{i}\left(  z\right)
dz}|0\rangle}{\langle0|e^{i\int\mathcal{L}\left(  z\right)  dz}|0\rangle}%
\nonumber \\
&=&\int d\vec{z}_{1}d\vec{z}_{2}d\vec{z}_{3}\,F_{1}\left(  p_{q},p_{\bar{q}}%
,p_{g},z_{0},\vec{z}_{1},\vec{z}_{2},\vec{z}_{3}\right)  ^{\alpha}%
\tr(U_{1}t^{a}U_{2}^{\dag}t^{b})U_{3}^{ba}%
\nonumber \\
&+&\int d\vec{z}_{1}d\vec{z}_{2}\,\tilde{F}_{2}\left(  p_{q},p_{\bar{q}}%
,p_{g},z_{0},\vec{z}_{1},\vec{z}_{2}\right)  ^{\alpha}\frac{N_{c}^{2}%
-1}{2N_{c}}\tr(U_{1}U_{2}^{\dag})\,.
\label{defMtilde}
\eqa
Here $(t^{b})_{l}^{k}$ is the projector to the color singlet state, and $c,$
$a,$ and $b$ are the gluon, quark and antiquark annihilation operators;
$F_{1}$ describes the contribution of the first two  diagrams and $\tilde
{F}_{2}$ stands for diagrams 3 and 4. The space coordinates $z_{0,1,2,3,4}$
and the momenta $p_{q,\bar{q},g}$\ are defined in fig.~\ref{3jetif}.

In this form $\tilde{M}$ contains contributions from terms without interaction in which
all operators $U$ are reduced to identity. To get the impact factor we have to subtract those terms. This amounts in replacing $\tilde{M}$ by $M\,,$ which reads
\beqa
&&M^{\alpha}=\int d\vec{z}_{1}d\vec{z}_{2}d\vec{z}_{3}\, F_{1}\left(  p_{q}%
,p_{\bar{q}},p_{g},z_{0},\vec{z}_{1},\vec{z}_{2},\vec{z}_{3}\right)  ^{\alpha
}\left[  \tr(U_{1}t^{a}U_{2}^{\dag}t^{b})\, U_{3}^{ba}-\frac{N_{c}^{2}-1}%
{2}\right]
\nonumber \\
&&
+\int d\vec{z}_{1}d\vec{z}_{2}\,\tilde{F}_{2}\left(  p_{q},p_{\bar{q}}%
,p_{g},z_{0},\vec{z}_{1},\vec{z}_{2}\right)  ^{\alpha}\frac{N_{c}^{2}%
-1}{2N_{c}}\left(  \tr(U_{1}U_{2}^{\dag})-N_{c}\right)  \label{M}%
\\
&&
=\int d\vec{z}_{1}d\vec{z}_{2}d\vec{z}_{3} \, F_{1}\left(  p_{q},p_{\bar{q}}%
,p_{g},z_{0},\vec{z}_{1},\vec{z}_{2},\vec{z}_{3}\right)  ^{\alpha}\frac{1}%
{2}\left(  \tr(U_{1}U_{3}^{\dag})\,\tr(U_{3}U_{2}^{\dag})-N_{c}\,\tr(U_{1}U_{2}%
^{\dag})\right)
\nonumber \\
&&
+\int d\vec{z}_{1}d\vec{z}_{2} \, F_{2}\left(  p_{q},p_{\bar{q}},p_{g},z_{0}%
,\vec{z}_{1},\vec{z}_{2}\right)  ^{\alpha}\frac{N_{c}^{2}-1}{2N_{c}}\left(
\tr(U_{1}U_{2}^{\dag})-N_{c}\right)  . \label{F2tilde}%
\eqa
Here%
\beqa
\hspace{-.4cm}F_{2}\left(  p_{q},p_{\bar{q}},p_{g},z_{0},\vec{z}_{1},\vec{z}_{2}\right)
^{\alpha}\!=\!\tilde{F}_{2}\left(  p_{q},p_{\bar{q}},p_{g},z_{0},\vec{z}_{1}%
,\vec{z}_{2}\right)  ^{\alpha}\!+\!\int d\vec{z}_{3}\,F_{1}\left(  p_{q},p_{\bar{q}%
},p_{g},z_{0},\vec{z}_{1},\vec{z}_{2},\vec{z}_{3}\right)  ^{\alpha}\!,
\label{F2+intF3}
\eqa
and
\beqa
\label{structureU}
\tr(U_{1}t^{a}U_{2}^{\dag}t^{b})U_{3}^{ba}=\frac{1}{2}\left(  \tr(U_{1}%
U_{3}^{\dag})\,\tr(U_{3}U_{2}^{\dag})-N_{c}\,\tr(U_{1}U_{2}^{\dag})\right)
+\frac{N_{c}^{2}-1}{2N_{c}}\tr(U_{1}U_{2}^{\dag})\,.
\eqa
The functions $F_{1}$ and $\tilde{F}_{2}$ are the two above mentioned components of the impact factor, which
we will calculate in the next two sections.

\section{Diagrams with the gluon crossing the shockwave}
\label{Sec:qqbarg-through-sw}

Using eqs.~(\ref{uinSW}--\ref{einSW}), the sum of the first two diagrams in fig.~\ref{3jetif}\ can be represented in the following way%
\begin{equation}
F_{1}\left(  z_{q},z_{\bar{q}},p_{g},z_{0},\vec{z}_{1},\vec{z}_{2},\vec{z}%
_{3}\right)  ^{\alpha}=\frac{ig}{(2\pi)^{2}}\int dz_{1}^{+}dz_{1}^{-}%
dz_{2}^{+}dz_{2}^{-}dz_{3}^{+}dz_{3}^{-}\delta(z_{1}^{+})\delta(z_{2}%
^{+})\delta(z_{3}^{+})L_{i\beta}^{j\alpha}R_{j}^{i\beta},
\end{equation}
where%
\beqa
\label{defL}
L_{i\beta}^{j\alpha}&=&\int dz_{4}\frac{1}{\frac{\partial\,\,\,\,}{\partial
z_{3}^{-}}}\frac{g_{\bot\beta\nu}z_{34}^{+}-z_{34}{}_{\bot\beta}n_{2\nu}%
}{\left(  z_{34}{}^{2}-i0\right)  ^{2}}%
\nonumber \\
&\times&\left[  \gamma^{+}\left(  G_{0}(z_{10})\gamma^{\alpha}G_{0}%
(z_{04})\gamma^{\nu}G_{0}(z_{42})+G_{0}(z_{14})\gamma^{\nu}G_{0}(z_{40}%
)\gamma^{\alpha}G_{0}(z_{02})\right)  \gamma^{+}\right]  _{i}^{j}\,,
\eqa
and
\begin{equation}
\label{defR}
R_{j}^{i\beta}=p_{g}^{+}\theta(p_{g}^{+})\frac{\varepsilon_{g}^{\ast\beta
}e^{ip_{g} \cdot z_{3}+ip_{q} \cdot z_{1}+ip_{_{\bar{q}}} \cdot z_{_{2}}}}{\sqrt{2p_{g}^{+}%
2p_{q}^{+}2p_{_{\bar{q}}}^{+}}}\left[  \gamma^{-}\gamma^{+}v_{p_{_{\bar{q}}}%
}\otimes\bar{u}_{p_{q}}\gamma^{+}\gamma^{-}\right]  _{j}^{i}.
\end{equation}
Using the integral (\ref{vertexIntegral}) and the fact that $z_{1,2,3}^{+}=0$ one
can write%
\beqa
\label{expL}
L_{i\beta}^{j\alpha}&=&\frac{2z_{30}^{+}}{\left(  2\pi\right)  ^{4}}\frac
{1}{\frac{\partial\,\,\,\,}{\partial z_{3}^{-}}}\left[  \gamma^{+}\left(
\frac{\hat{z}_{10}\gamma^{\alpha}\hat{z}_{30}\gamma_{\bot\beta}\hat{z}_{32}%
}{z_{30}^{4}z_{32}^{2}z_{20}{}^{2}z_{10}^{4}}+\frac{\hat{z}_{31}\gamma
_{\bot\beta}\hat{z}_{30}\gamma^{\alpha}\hat{z}_{02}}{z_{31}^{2}z_{30}%
^{4}z_{01}{}^{2}z_{02}^{4}}\right.  \right.
\nonumber \\
  &+&\left.  \left.\frac{2}{z_{30}^{2}}\frac{\hat{z}_{10}\gamma^{\alpha}\hat
{z}_{02}}{z_{20}{}^{4}z_{10}^{4}}\left(  \frac{z_{31\bot\beta}}{z_{31}^{2}%
}-\frac{z_{32\bot\beta}}{z_{32}^{2}}\right)  \right)  \gamma^{+}\right]
_{i}^{j}.
\eqa
Integrating w.r.t. $z_{3}^{-}$ with the help of eq.~(\ref{1/d}) we have%
\beqa
\label{int}
L_{i\beta}^{j\alpha}&=&\frac{-1}{\left(  2\pi\right)  ^{4}}\int_{\sigma
}^{+\infty}\frac{dp^{+}}{p^{+}}e^{-ip^{+}\left(  z_{30}^{-}-\frac{\vec{z}%
_{30}^{\,\,2}+i0}{2z_{30}^{+}}\right)  }\left[  \gamma^{+}\left(  \frac
{ip^{+}\hat{z}_{10}\gamma^{\alpha}(\gamma^{-}z_{30}^{+}+\hat{z}_{30\bot
})\gamma_{\bot\beta}\hat{z}_{32}}{(2z_{30}^{+})z_{32}^{2}z_{20}{}^{2}%
z_{10}^{4}}\right.  \right.
\nonumber \\
&+& \left.  \left.  \frac{ip^{+}\hat{z}_{31}\gamma_{\bot\beta}(\gamma^{-}%
z_{30}^{+}+\hat{z}_{30\bot})\gamma^{\alpha}\hat{z}_{02}}{z_{31}^{2}%
(2z_{30}^{+})z_{01}{}^{2}z_{02}^{4}}+2\frac{\hat{z}_{10}\gamma^{\alpha}\hat
{z}_{02}}{z_{20}{}^{4}z_{10}^{4}}\left(  \frac{z_{31\bot\beta}}{z_{31}^{2}%
}-\frac{z_{32\bot\beta}}{z_{32}^{2}}\right)  \right)  \gamma^{+}\right]
_{i}^{j},
\eqa
where $\sigma=e^{\eta}$ is the longitudinal cutoff~(\ref{cutoff}). As a result,%
\beqa
\label{resF1}
&&F_{1}\left(  p_{q},p_{\bar{q}},p_{g},z_{0},\vec{z}_{1},\vec{z}_{2},\vec{z}%
_{3}\right)  ^{\alpha}=\theta(p_{g}^{+}-\sigma)\frac{ig}{(2\pi)^{3}%
}\varepsilon_{g}^{\ast\beta}\,\frac{e^{iz_{0}^{-}(p_{q}^{+}+p_{_{\bar{q}}}%
^{+}+p_{g}^{+})-i\vec{p}_{q} \cdot \vec{z}_{1}-i\vec{p}_{_{\bar{q}}} \cdot \vec{z}_{_{2}%
}-i\vec{p}_{g} \cdot \vec{z}_{3}}}{(2z_{0}^{+})^{2}\sqrt{2p_{q}^{+}2p_{_{\bar{q}}%
}^{+}2p_{g}^{+}}}%
\nonumber \\
&\times&\bar{u}_{p_{q}}\gamma^{+}\left[  \frac{z_{32\bot}^{\omega}}%
{-\vec{z}_{32}^{\,\,2}}\left(ip_{q}^{+}\gamma^{-}-\gamma_{\bot}^{\sigma}%
\frac{\partial}{\partial z_{1\bot}^{\sigma}}\right)\gamma^{\alpha}\left(ip_{g}^{+}%
\gamma^{-}-\gamma_{\bot}^{\rho}\frac{\partial}{\partial z_{3\bot}^{\rho}%
}\right)\gamma_{\bot\beta}\gamma_{\bot\omega}\right.
\nonumber \\
&+&\frac{\hat{z}_{31\bot}}{\vec{z}_{31}^{\,\,2}}\gamma_{\bot\beta}\left(ip_{g}%
^{+}\gamma^{-}-\gamma_{\bot}^{\rho}\frac{\partial}{\partial z_{3\bot}^{\rho}%
}\right)\gamma^{\alpha}\left(ip_{_{\bar{q}}}^{+}\gamma^{-}-\gamma_{\bot}^{\sigma}%
\frac{\partial}{\partial z_{2\bot}^{\sigma}}\right)
\nonumber \\
&+& \left.  2\left(  \frac{z_{31\bot\beta}}{\vec{z}_{31}^{\,\,2}}-\frac
{z_{32\bot\beta}}{\vec{z}_{32}^{\,\,2}}\right)  \left(ip_{q}^{+}\gamma^{-}%
-\gamma_{\bot}^{\rho}\frac{\partial}{\partial z_{1\bot}^{\rho}}\right)\gamma
^{\alpha}\left(ip_{_{\bar{q}}}^{+}\gamma^{-}-\gamma_{\bot}^{\sigma}\frac{\partial
}{\partial z_{2\bot}^{\sigma}}\right)\right]  \gamma^{+}v_{p_{_{\bar{q}}}} \nonumber \\
&\times& e^{-i\frac{p_{g}^{+}\vec{z}_{30}^{\,\,2}+p_{q}^{+}\vec{z}_{10}^{\,\,2}%
+p_{_{\bar{q}}}^{+}\vec{z}_{20}^{\,\,2}+i0}{2z_{0}^{+}}}.
\eqa
Via eq.~(\ref{Kint}) we will calculate the Fourier transform of $F_{1}$
\begin{equation}
F_{1}\left(  p_{q},p_{\bar{q}},p_{g},k,\vec{z}_{1},\vec{z}_{2},\vec{z}%
_{3}\right)  ^{\alpha}=\int d^{4}z_{0}\, e^{-ik \cdot z_{0}}F_{1}\left(  p_{q}%
,p_{\bar{q}},p_{g},z_{0},\vec{z}_{1},\vec{z}_{2},\vec{z}_{3}\right)  ^{\alpha}%
\end{equation}
for the photon in our kinematics (\ref{photonk}).  Using our notation (\ref{xs}) and denoting
\begin{equation}
Z_{123}=\sqrt{x_{g}x_{q}\vec{z}_{13}^{\,\,2}+x_{q}x_{_{\bar{q}}}\vec{z}%
_{21}^{\,\,2}+x_{_{\bar{q}}}x_{g}\vec{z}_{32}^{\,\,2}}\,,
\label{Z123}%
\end{equation}
we get%
\beqa
\label{resF1Fourier}
&&\hspace{-.5cm}F_{1}\left(  p_{q},p_{\bar{q}},p_{g},k,\vec{z}_{1},\vec{z}_{2},\vec{z}%
_{3}\right)  ^{\alpha}=-\frac{g\, \theta(p_{g}^{+}-\sigma)}{4\pi k^{+}}%
\delta(k^{+}-p_{g}^{+}-p_{q}^{+}-p_{_{\bar{q}}}^{+})\, \varepsilon_{g\beta}%
^{\ast}\,\frac{e^{-i\vec{p}_{q} \cdot \vec{z}_{1}-i\vec{p}_{_{\bar{q}}} \cdot \vec{z}_{_{2}%
}-i\vec{p}_{g} \cdot \vec{z}_{3}}}{\sqrt{2p_{q}^{+}2p_{_{\bar{q}}}^{+}2p_{g}^{+}}}%
\nonumber \\
&\times&\!\bar{u}_{p_{q}}\gamma^{+}\left[  \frac{z_{32\bot}^{\omega}}%
{-\vec{z}_{32}^{\,\,2}}\left(ip_{q}^{+}\gamma^{-}-\gamma_{\bot}^{\sigma}%
\frac{\partial}{\partial z_{1\bot}^{\sigma}}\right)\gamma^{\alpha}\left(ip_{g}^{+}%
\gamma^{-}-\gamma_{\bot}^{\rho}\frac{\partial}{\partial z_{3\bot}^{\rho}%
}\right)\gamma_{\bot}^{\beta}\gamma_{\bot\omega}\right.
\nonumber \\
&+&\!\frac{\hat{z}_{31\bot}}{\vec{z}_{31}^{\,\,2}}\gamma_{\bot}^{\beta}\left(ip_{g}%
^{+}\gamma^{-}-\gamma_{\bot}^{\rho}\frac{\partial}{\partial z_{3\bot}^{\rho}%
}\right)\gamma^{\alpha}\left(ip_{_{\bar{q}}}^{+}\gamma^{-}-\gamma_{\bot}^{\sigma}%
\frac{\partial}{\partial z_{2\bot}^{\sigma}}\right)
\nonumber \\
&+& \!\left.  2\left(  \frac{z_{31\bot}^{\beta}}{\vec{z}_{31}^{\,\,2}}%
-\frac{z_{32\bot}^{\beta}}{\vec{z}_{32}^{\,\,2}}\right)  \left(ip_{q}^{+}\gamma
^{-}-\gamma_{\bot}^{\rho}\frac{\partial}{\partial z_{1\bot}^{\rho}}%
\right)\gamma^{\alpha}\left(ip_{_{\bar{q}}}^{+}\gamma^{-}-\gamma_{\bot}^{\sigma}%
\frac{\partial}{\partial z_{2\bot}^{\sigma}}\right)\right]  \gamma^{+}%
v_{p_{_{\bar{q}}}}K_{0}(QZ_{123}).\
\eqa
As before for the $\gamma\rightarrow q\bar{q}$ impact factor, we encounter contributions involving  $\delta$ distributions
  for $\alpha=-.$ Indeed, e.g.%
\begin{equation}
\frac{\partial^{2}K_{0}(QZ_{123})}{\partial z_{1\bot}^{\sigma}\partial
z_{3\bot}^{\rho}}=Q^{2}\frac{\left(  QZ_{123}K_{0}^{\prime}(QZ_{123})\right)
^{\prime}}{QZ_{123}}\frac{\partial Z_{123}}{\partial z_{1\bot}^{\sigma}}%
\frac{\partial Z_{123}}{\partial z_{3\bot}^{\rho}}+QZ_{123}K_{0}^{\prime
}(QZ_{123})\frac{\partial^{2}\ln Z_{123}}{\partial z_{1\bot}^{\sigma}\partial
z_{3\bot}^{\rho}},
\end{equation}
and from eq.~(\ref{Klaplasian}) we obtain
\begin{equation}
\frac{\left(  QZ_{123}K_{0}^{\prime}(QZ_{123})\right)  ^{\prime}}{QZ_{123}%
}=K_{0}(QZ_{123})+4\pi\delta\left(  Q^{2}Z_{123}^{2}\right)  \delta\left(
\phi\right)  .
\end{equation}
Again the term with the delta function gives a vanishing contribution to $M$ introduced in eq.~(\ref{M}) and we drop it. Then, using eq.~(\ref{Klaplasian}) and the matrix elements
(\ref{ug+v}), we get%
\begin{equation}
F_{1}^{+}=F_{1}^{-}\frac{2(k^{+})^{2}}{Q^{2}}\,.
\end{equation}
On one hand this implies the electromagnetic gauge invariance
\begin{equation}
F_{1}\left(  p_{q},p_{\bar{q}},p_{g},k,\vec{z}_{1},\vec{z}_{2},\vec{z}%
_{3}\right)  ^{\alpha}k_{\alpha}=0,
\end{equation}
and on the second hand it gives the contribution to the impact factor for longitudinal photon (\ref{eL})%
\beqa
\label{resF1L}
&&F_{1}\left(  p_{q},p_{\bar{q}},p_{g},k,\vec{z}_{1},\vec{z}_{2},\vec{z}%
_{3}\right)  ^{\alpha}\varepsilon_{L\alpha}  \\
&=&-\delta(k^{+}-p_{g}^{+}-p_{q}%
^{+}-p_{_{\bar{q}}}^{+})\theta(p_{g}^{+}-\sigma)Qg\frac{e^{-i\vec{p}_{q} \cdot %
\vec{z}_{1}-i\vec{p}_{_{\bar{q}}} \cdot \vec{z}_{_{2}}-i\vec{p}_{g} \cdot \vec{z}_{3}}}%
{\pi\sqrt{2p_{g}^{+}}}K_{0}(QZ_{123})
\nonumber \\
&\times&\delta_{\lambda_{q},-\lambda_{\bar{q}}}\varepsilon_{g}^{\ast\beta
}\left(  i\lambda_{q}\epsilon^{\gamma\beta3}x_{g}\left\{  x_{_{\bar{q}}}%
\frac{z_{31}^{\gamma}}{\vec{z}_{31}^{\,\,2}}+x_{q}\frac{z_{32}^{\gamma}}%
{\vec{z}_{32}^{\,\,2}}\right\}  +\left\{  (2x_{q}+x_{g})x_{_{\bar{q}}}%
\frac{z_{31}^{\beta}}{\vec{z}_{31}^{\,\,2}}-(2x_{_{\bar{q}}}+x_{g})x_{q}%
\frac{z_{32}^{\beta}}{\vec{z}_{32}^{\,\,2}}\right\}  \right)  . \nonumber
\eqa
Expressing the vector products $e^{\gamma\beta3}a^{\gamma}b^{\beta}=[\vec
{a}\times\vec{b}]$ through the scalar products using eq.~(\ref{crossToSP}) we have%
\beqa
&&\hspace{-1cm}F_{1}\left(  p_{q},p_{\bar{q}},p_{g},k,\vec{z}_{1},\vec{z}_{2},\vec{z}%
_{3}\right)  ^{\alpha}\varepsilon_{L\alpha}\nonumber \\
&=&\delta(k^{+}-p_{g}^{+}-p_{q}%
^{+}-p_{_{\bar{q}}}^{+})\theta(p_{g}^{+}-\sigma)2Qg\frac{e^{-i\vec{p}_{q} \cdot %
\vec{z}_{1}-i\vec{p}_{_{\bar{q}}} \cdot \vec{z}_{_{2}}-i\vec{p}_{g} \cdot \vec{z}_{3}}}%
{\pi\sqrt{2p_{g}^{+}}}K_{0}(QZ_{123})
\nonumber \\
&\times&\delta_{\lambda_{q},-\lambda_{\bar{q}}}\left\{  (x_{_{\bar{q}}}%
+x_{g}\delta_{-s_{g}\lambda_{q}})x_{q}\frac{\vec{z}_{32} \cdot \vec{\varepsilon}%
_{g}^{\,\,\ast}}{\vec{z}_{32}^{\,\,2}}-(x_{q}+x_{g}\delta_{-s_{g}%
\lambda_{\bar{q}}})x_{_{\bar{q}}}\frac{\vec{z}_{31} \cdot \vec{\varepsilon}%
_{g}^{\,\,\ast}}{\vec{z}_{31}^{\,\,2}}\right\}  .\label{F1eL}%
\eqa
For transverse photon (\ref{eT}) we get the following contribution to the
impact factor%
\beqa
&&\hspace{-.7cm}F_{1}\left(  p_{q},p_{\bar{q}},p_{g},k,\vec{z}_{1},\vec{z}_{2},\vec{z}%
_{3}\right)  ^{\alpha}\varepsilon_{T\alpha}=2igQ\delta(k^{+}-p_{g}^{+}%
-p_{q}^{+}-p_{_{\bar{q}}}^{+})\theta(p_{g}^{+}-\sigma)
\frac{e^{-i\vec{p}_{q} \cdot \vec{z}_{1}-i\vec{p}_{_{\bar{q}}} \cdot \vec{z}_{_{2}%
}-i\vec{p}_{g} \cdot \vec{z}_{3}}}{\pi Z_{123}\sqrt{2p_{g}^{+}}}
\nonumber \\
&\times&\delta_{\lambda
_{q},-\lambda_{\bar{q}}}K_{1}(QZ_{123})\left\{  -\frac{\left(  \vec{z}%
_{23} \cdot \vec{\varepsilon}_{g}^{\,\,\ast}\right)  \left(  \vec{z}_{13} \cdot %
\vec{\varepsilon}_{T}\right)  }{\vec{z}_{23}{}^{2}}x_{q}\left(  x_{q}%
-\delta_{s\lambda_{\bar{q}}}\right)  \left(  x_{\bar{q}}+x_{g}\delta
_{-s_{g}\lambda_{q}}\right)  \right.
\nonumber \\
&&
\left.  -\frac{\left(  \vec{z}_{23} \cdot \vec{\varepsilon}_{g}^{\,\,\ast}\right)
\left(  \vec{z}_{23} \cdot \vec{\varepsilon}_{T}\right)  }{\vec{z}_{23}{}^{2}}%
x_{q}x_{\bar{q}}\left(  x_{\bar{q}}+x_{g}\delta_{-s_{g}\lambda_{q}}%
-\delta_{s\lambda_{q}}\right)  \right\}  -\left(  q\leftrightarrow\bar
{q}\right)  , \label{F1eT}%
\eqa
where%
\begin{equation}
\left(  q\leftrightarrow\bar{q}\right)  \equiv\left(  \lambda_{q},x_{q}%
,\vec{z}_{1},\vec{p}_{q}\leftrightarrow\lambda_{\bar{q}},x_{\bar{q}},\vec
{z}_{2},\vec{p}_{\bar{q}}\right)  .
\end{equation}
One can check that the results (\ref{F1eL}) and (\ref{F1eT}) are compatible with the wave function derived in ref.~\cite{Beuf:2011xd}.

\section{Diagrams without the gluon crossing the shockwave}
\label{Sec:qqbarg-no-sw}

Here we will calculate $\tilde{F}_{2}$ (\ref{F2tilde}), which gives the
contribution from diagrams 3 and 4 in fig.~\ref{3jetif}. It reads
\begin{equation}
\tilde{F}_{2}\left(  z_{q},z_{\bar{q}},p_{g},z_{0},\vec{z}_{1},\vec{z}%
_{2}\right)  ^{\alpha}=\frac{ig}{4}\int dz_{1}^{+}dz_{1}^{-}dz_{2}^{+}%
dz_{2}^{-}\delta\left(  z_{1}^{+}\right)  \delta\left(  z_{2}^{+}\right)
L_{i}^{\prime j\alpha}R_{j}^{\prime i}
\end{equation}%
where
\begin{equation}
\label{defL'}
L_{i}^{\prime j\alpha}=\left[  \gamma^{+}G_{0}(z_{10})\gamma^{\alpha}%
G_{0}(z_{02})\gamma^{+}\right]  _{i}^{j}\,,
\end{equation}%
and
\beqa
\label{defR'}
&&R_{j}^{\prime i\beta}=\int\theta\left(  z_{4}^{+}\right)  d^{4}z_{4}%
\frac{\varepsilon_{g}^{\ast\beta}e^{ip_{g} \cdot z_{4}}}{\sqrt{2p_{g}^{+}2p_{q}%
^{+}2p_{_{\bar{q}}}^{+}}}\nonumber \\
&&\hspace{-1cm}\times\left[  \gamma^{-}\gamma^{+}\!\left(  e^{ip_{q} \cdot z_{1}+ip_{_{\bar{q}}} \cdot z_{4}%
}G_{0}(z_{24})\gamma_{\beta}v_{p_{_{\bar{q}}}}\otimes\bar{u}_{p_{q}%
}+e^{ip_{q} \cdot z_{4}+ip_{_{\bar{q}}} \cdot z_{_{2}}}v_{p_{_{\bar{q}}}}\otimes\overline
{u}_{p_{q}}\gamma_{\beta}G_{0}(z_{41})\right)  \gamma^{+}\gamma^{-}\!\right]
_{j}^{i}.
\eqa
Integrating eq.~(\ref{defR'}) w.r.t. $z_{4},$ we get%
\beqa
\label{intR'}
&&\hspace{-.25cm}R_{j}^{\prime i}=-i\frac{\varepsilon_{g}^{\ast\beta}}{\sqrt{2p_{g}^{+}%
2p_{q}^{+}2p_{_{\bar{q}}}^{+}}}e^{ip_{q}^{+}z_{1}^{-}-i\vec{p}_{q} \cdot \vec{z}%
_{1}+ip_{_{\bar{q}}}^{+}z_{2}^{-}-i\vec{p}_{_{\bar{q}}} \cdot \vec{z}_{2}}\!\left[
\gamma^{-}\gamma^{+}\!\left( \! e^{ip_{g}^{+}z_{2}^{-}-i\vec{p}_{g} \cdot \vec{z}_{2}%
}\frac{(\hat{p}_{_{\bar{q}}}+\hat{p}_{g})}{(p_{\bar{q}}+p_{g})^{2}}%
\gamma_{\beta}v_{p_{_{\bar{q}}}}\otimes\bar{u}_{p_{q}}\right.  \right.\!\!\!
\nonumber \\
&&\left.  \left.  -v_{p_{_{\bar{q}}}}\otimes\bar{u}_{p_{q}}\gamma_{\beta
}\frac{(\hat{p}_{q}+\hat{p}_{g})}{(p_{q}+p_{g})^{2}}e^{ip_{g}^{+}z_{1}%
^{-}-i\vec{p}_{g} \cdot \vec{z}_{1}}\right)  \gamma^{+}\gamma^{-}\right]  _{j}%
^{i}\,\,. \!\!\!
\eqa
As a result,%
\beqa
\label{resF2tilde}
&&\hspace{-.2cm}\tilde{F}_{2}\left(  p_{q},p_{\bar{q}},p_{g},z_{0},\vec{z}_{1},\vec{z}%
_{2}\right)  ^{\alpha}=\theta(p_{g}^{+}-\sigma)\frac{g}{16\pi^{2}}%
\frac{\varepsilon_{g}^{\ast\beta}\theta(p_{g}^{+})}{\sqrt{2p_{g}^{+}2p_{q}%
^{+}2p_{_{\bar{q}}}^{+}}}\frac{1}{\left(  z_{0}^{+}\right)  ^{2}%
}e^{i(p_{_{\bar{q}}}^{+}+p_{q}^{+}+p_{g}^{+})z_{0}^{-}-i\vec{p}_{q} \cdot \vec{z}%
_{1}-i\vec{p}_{_{\bar{q}}} \cdot \vec{z}_{2}}%
\nonumber \\
&&\hspace{-.2cm}\times\left\{  -e^{-i\vec{p}_{g} \cdot \vec{z}_{2}}\bar{u}_{p_{q}}\gamma
^{+}\left(  ip_{q}^{+}\gamma^{-}-\gamma_{\bot}^{\sigma}\frac{\partial
}{\partial z_{1\bot}^{\sigma}}\right)  \gamma^{\alpha}\left(  i(p_{g}%
^{+}+p_{_{\bar{q}}}^{+})\gamma^{-}-\gamma_{\bot}^{\sigma}\frac{\partial
}{\partial z_{2\bot}^{\sigma}}\right)  \gamma^{+}\frac{(\hat{p}_{_{\bar{q}}%
}+\hat{p}_{g})}{(p_{\bar{q}}+p_{g})^{2}}\gamma_{\beta}v_{p_{_{\bar{q}}}%
}\right.
\nonumber \\
&&\hspace{-.2cm}\times e^{-ik^{+}\frac{(x_{g}+x_{\bar{q}})\vec{z}_{20}^{\,\,2}+x_{q}\vec
{z}_{10}^{\,\,2}+i0}{2z_{0}^{+}}}+e^{-i\vec{p}_{g}\vec{z}_{1}}\overline
{u}_{p_{q}}\gamma_{\beta}\frac{(\hat{p}_{q}+\hat{p}_{g})}{(p_{q}+p_{g})^{2}%
}\gamma^{+}\left(  i(p_{g}^{+}+p_{q}^{+})\gamma^{-}-\gamma_{\bot}^{\sigma
}\frac{\partial}{\partial z_{1\bot}^{\sigma}}\right)
\nonumber \\
&&\hspace{-.2cm}\times\left.  \gamma^{\alpha}\left(  ip_{_{\bar{q}}}^{+}\gamma^{-}%
-\gamma_{\bot}^{\sigma}\frac{\partial}{\partial z_{2\bot}^{\sigma}}\right)
\gamma^{+}v_{p_{_{\bar{q}}}}e^{-ik^{+}\frac{x_{\bar{q}}\vec{z}_{20}%
^{\,\,2}+(x_{g}+x_{q})\vec{z}_{10}^{\,\,2}+i0}{2z_{0}^{+}}}\right\}  .
\eqa
Next we calculate the Fourier transform of $\tilde{F}_{2}$%
\begin{equation}
\tilde{F}_{2}\left(  p_{q},p_{\bar{q}},p_{g},k,\vec{z}_{1},\vec{z}_{2}\right)
^{\alpha}=\int d^{4}z_{0}\, e^{-ik \cdot z_{0}}\tilde{F}_{2}\left(  p_{q},p_{\bar{q}%
},p_{g},z_{0},\vec{z}_{1},\vec{z}_{2}\right)  ^{\alpha}%
\end{equation}
for the photon in our kinematics (\ref{photonk})%
\beqa
&&\tilde{F}_{2}\left(  p_{q},p_{\bar{q}},p_{g},k,\vec{z}_{1},\vec{z}_{2}\right)
^{\alpha}=\frac{ig}{2k^{+}}\theta(p_{g}^{+}-\sigma)\,\delta(k^{+}-p_{g}%
^{+}-p_{q}^{+}-p_{_{\bar{q}}}^{+})\, \varepsilon_{g}^{\ast\beta}\frac
{e^{-i\vec{p}_{q} \cdot \vec{z}_{1}-i\vec{p}_{_{\bar{q}}} \cdot \vec{z}_{2}}}{\sqrt
{2p_{g}^{+}2p_{q}^{+}2p_{_{\bar{q}}}^{+}}}%
\nonumber \\
&&\times\left[  -e^{-i\vec{p}_{g} \cdot \vec{z}_{2}}\bar{u}_{p_{q}}\gamma
^{+}\left(  ip_{q}^{+}\gamma^{-}-\gamma_{\bot}^{\sigma}\frac{\partial
}{\partial z_{1\bot}^{\sigma}}\right)  \gamma^{\alpha}\left(  i(p_{g}%
^{+}+p_{_{\bar{q}}}^{+})\gamma^{-}-\gamma_{\bot}^{\sigma}\frac{\partial
}{\partial z_{2\bot}^{\sigma}}\right)  \gamma^{+}\right.
\nonumber \\
&&\times\frac{(\hat{p}_{_{\bar{q}}}+\hat{p}_{g})}{(p_{\bar{q}}+p_{g})^{2}}%
\gamma_{\beta}v_{p_{_{\bar{q}}}}K_{0}(QZ_{122})+e^{-i\vec{p}_{g}\cdot\vec{z}_{1}%
}\bar{u}_{p_{q}}\gamma_{\beta}\frac{(\hat{p}_{q}+\hat{p}_{g})}%
{(p_{q}+p_{g})^{2}}%
\nonumber \\
&&\times\left.  \gamma^{+}\left(  i(p_{g}^{+}+p_{q}^{+})\gamma^{-}-\gamma_{\bot
}^{\sigma}\frac{\partial}{\partial z_{1\bot}^{\sigma}}\right)  \gamma^{\alpha
}\left(  ip_{_{\bar{q}}}^{+}\gamma^{-}-\gamma_{\bot}^{\sigma}\frac{\partial
}{\partial z_{2\bot}^{\sigma}}\right)  \gamma^{+}v_{p_{_{\bar{q}}}}%
K_{0}(QZ_{121})\right]  .
\eqa
In the arguments of the McDonald functions we encounter the following structures%
\begin{equation}
Z_{122}=Z_{123}|_{z_{3}\rightarrow z_{2}}=\sqrt{x_{q}\left(  x_{g}%
+x_{_{\bar{q}}}\right)  \vec{z}_{12}^{\,\,2}}=\sqrt{x_{q}\left(
1-x_{q}\right)  \vec{z}_{12}^{\,\,2}},
\end{equation}%
\begin{equation}
Z_{121}=Z_{123}|_{z_{3}\rightarrow z_{1}}=\sqrt{\left(  x_{q}+x_{g}\right)
x_{_{\bar{q}}}\vec{z}_{21}^{\,\,2}}=\sqrt{\left(  1-x_{\bar{q}}\right)
x_{_{\bar{q}}}\vec{z}_{21}^{\,\,2}}.
\end{equation}
Again using the Bessel equation (\ref{Klaplasian}) as well as the matrix elements (\ref{ug+v}) and
(\ref{ugjv}), then dropping the corresponding $\delta$ distributions  we can check the
electromagnetic gauge invariance
\begin{equation}
\tilde{F}_{2}\left(  p_{q},p_{\bar{q}},p_{g},k,\vec{z}_{1},\vec{z}_{2}\right)
^{\alpha}k_{\alpha}=0.
\end{equation}
Then taking into account Dirac equation and the gauge condition
(\ref{guageProector}), we get the contribution to the impact factor for
longitudinal photon (\ref{eL})%
\beqa
\label{resF2L}
&&\tilde{F}_{2}\left(  p_{q},p_{\bar{q}},p_{g},k,\vec{z}_{1},\vec{z}_{2}\right)
^{\alpha}\varepsilon_{L\alpha}=4ig \, Q\,\theta(p_{g}^{+}-\sigma)\delta(k^{+}%
-p_{g}^{+}-p_{q}^{+}-p_{_{\bar{q}}}^{+})\frac{e^{-i\vec{p}_{q} \cdot \vec{z}%
_{1}-i\vec{p}_{_{\bar{q}}} \cdot \vec{z}_{2}}}{\sqrt{2p_{g}^{+}}}%
\nonumber \\
&&\times\delta_{\lambda_{q},-\lambda_{\bar{q}}}\frac{x_{q}\left(  x_{g}%
+x_{\bar{q}}\right)  \left(  \delta_{-s_{g}\lambda_{q}}x_{g}+x_{\bar{q}%
}\right)  }{\left(  p_{g}+p_{\bar{q}}\right)  {}^{2}}(\vec{P}_{\bar{q}}{} \cdot %
\vec{\varepsilon}_{g}^{\,\,\ast})\,e^{-i\vec{p}_{g} \cdot \vec{z}_{2}}K_{0}%
(QZ_{122})-\left(  q\leftrightarrow\bar{q}\right)  ,
\eqa
where $P_{q,\bar{q}}$ are defined in eq.~(\ref{PP}). For transverse photons
(\ref{eT}) we have
\beqa
\label{resF2tildeT}
&&\tilde{F}_{2}\left(  p_{q},p_{\bar{q}},p_{g},k,\vec{z}_{1},\vec{z}_{2}\right)
^{\alpha}\varepsilon_{T\alpha}=-4g\,Q\,\theta(p_{g}^{+}-\sigma)\,\delta(k^{+}%
-p_{g}^{+}-p_{q}^{+}-p_{_{\bar{q}}}^{+})\frac{e^{-i\vec{p}_{q} \cdot \vec{z}%
_{1}-i\vec{p}_{_{\bar{q}}} \cdot \vec{z}_{2}}}{\sqrt{2p_{g}^{+}}}\delta_{\lambda
_{q},-\lambda_{\bar{q}}}%
\nonumber \\
&&\times\left(  \vec{z}_{12} \cdot \vec{\varepsilon}_{T}\right)  \left(  \delta
_{\lambda_{\bar{q}}s}-x_{q}\right)  \left(  \delta_{-s_{g}\lambda_{q}}%
x_{g}+x_{\bar{q}}\right)  (\vec{P}_{\bar{q}} \cdot \vec{\varepsilon}_{g}^{\,\,\ast
})\frac{x_{q}\left(  x_{g}+x_{\bar{q}}\right)  }{\left(  p_{g}+p_{\bar{q}%
}\right)  {}^{2}}\frac{K_{1}(QZ_{122})}{Z_{122}}e^{-i\vec{p}_{g} \cdot \vec{z}_{2}%
}-\left(  q\leftrightarrow\bar{q}\right)  .\nonumber \\
\eqa

\section{Impact factor for interaction with color dipole}
\label{Sec:dipole}

In the 2- and 3-gluon approximations (BFKL and BKP) of exchanges in $t$-channel one needs the Green
function obeying the linear equation. In the color singlet channel, the subtracted color
dipole is the operator that plays this role
\begin{equation}
\mathbf{U}_{12}=\frac{1}{N_{c}}\tr\left(  U_{1}U_{2}^{\dag}\right)  -1\,.
\end{equation}%
The operator  appearing in eq.~(\ref{F2tilde})
can be rewritten as
\begin{equation}
\tr(U_{1}U_{3}^{\dag})\, \tr(U_{3}U_{2}^{\dag})-N_{c}\,\tr(U_{1}U_{2}^{\dag}%
)=N_{c}^{2}\left(  \mathbf{U}_{32}+\mathbf{U}_{13}-\mathbf{U}_{12}%
+\mathbf{U}_{32}\mathbf{U}_{13}\right)  .
\end{equation}
Therefore in the 2- and 3-gluon approximations in which we neglect $\mathbf{U}%
_{32}\mathbf{U}_{13}\,,$ eq.~(\ref{M}) reads
\beqa
\label{M3g}
&&\hspace{-.3cm}M^{\alpha}\!\overset{\mathrm{g^3}}{=}\!\!\int d\vec{z}_{1}d\vec{z}_{2}\mathbf{U}%
_{12}\!\left\{ \! \frac{N_{c}^{2}-1}{2}\tilde{F}_{2}\left(  p_{q},p_{\bar{q}%
},p_{g},z_{0},\vec{z}_{1},\vec{z}_{2}\right)  ^{\alpha}-\frac{1}{2}\int
d\vec{z}_{3}F_{1}\left(  p_{q},p_{\bar{q}},p_{g},z_{0},\vec{z}_{1},\vec{z}%
_{2},\vec{z}_{3}\right)  ^{\alpha}\!\right\}
\nonumber \\
&&+\frac{1}{2}N_{c}^{2}\int d\vec{z}_{1}d\vec{z}_{3}\mathbf{U}_{13}\int d\vec
{z}_{2}F_{1}\left(  p_{q},p_{\bar{q}},p_{g},z_{0},\vec{z}_{1},\vec{z}_{2}%
,\vec{z}_{3}\right)  ^{\alpha}%
\nonumber \\
&&+\frac{1}{2}N_{c}^{2}\int d\vec{z}_{2}d\vec{z}_{3}\mathbf{U}_{32}\int d\vec
{z}_{1}F_{1}\left(  p_{q},p_{\bar{q}},p_{g},z_{0},\vec{z}_{1},\vec{z}_{2}%
,\vec{z}_{3}\right)  ^{\alpha},%
\eqa
which we write as
\begin{equation}
\label{M3gBis}
M^{\alpha}\overset{\mathrm{g^3}}{=}\frac{1}{2}\int d\vec{z}_{1}d\vec{z}%
_{2}\mathbf{U}_{12}\left\{  \tilde{F}_{1}\left(  p_{q},p_{\bar{q}},p_{g}%
,z_{0},\vec{z}_{1},\vec{z}_{2}\right)  ^{\alpha}+\left(  N_{c}^{2}-1\right)
\tilde{F}_{2}\left(  p_{q},p_{\bar{q}},p_{g},z_{0},\vec{z}_{1},\vec{z}%
_{2}\right)  ^{\alpha}\right\}  \,,%
\end{equation}
where%
\beqa
&&\tilde{F}_{1}\left(  p_{q},p_{\bar{q}},p_{g},z_{0},\vec{z}_{1},\vec{z}%
_{2}\right)  ^{\alpha}=\int d\vec{z}_{3}\left[  N_{c}^{2}F_{1}\left(
p_{q},p_{\bar{q}},p_{g},z_{0},\vec{z}_{1},\vec{z}_{3},\vec{z}_{2}\right)
^{\alpha}\right.
\nonumber \\
&&+N_{c}^{2}F_{1}\left(  p_{q},p_{\bar{q}},p_{g},z_{0},\vec{z}_{3},\vec{z}%
_{2},\vec{z}_{1}\right)  ^{\alpha}-\left.  F_{1}\left(  p_{q},p_{\bar{q}%
},p_{g},z_{0},\vec{z}_{1},\vec{z}_{2},\vec{z}_{3}\right)  ^{\alpha}\right]
.\label{F1tilde}%
\eqa
The integrals required in order to calculate expression (\ref{F1tilde}) are discussed
in appendix A. Using (\ref{int_K0}) and (\ref{int_zK0}) for longitudinal
photon (\ref{eL}) we have
\beqa
\label{intF1L1}
&&\hspace{-.3cm}\int d\vec{z}_{3}F_{1}\left(  p_{q},p_{\bar{q}},p_{g},k,\vec{z}_{1},\vec
{z}_{2},\vec{z}_{3}\right)  ^{\alpha}\varepsilon_{L\alpha}=-\delta
_{\lambda_{q},-\lambda_{\bar{q}}}\delta(k^{+}-p_{g}^{+}-p_{q}^{+}-p_{_{\bar
{q}}}^{+})\theta(p_{g}^{+}-\sigma)
\nonumber \\
&&\hspace{-.3cm}\times\frac{2Qg}{\sqrt{2p_{g}^{+}}}\frac{e^{-i\vec{p}_{q} \cdot \vec{z}_{1}-i\vec
{p}_{_{\bar{q}}} \cdot \vec{z}_{_{2}}}}{(1-x_{g})x_{g}}(x_{_{\bar{q}}}+x_{g}%
\delta_{-s_{g}\lambda_{q}})x_{q}e^{-i\vec{p}_{g} \cdot \vec{z}_{2}}\int_{0}%
^{1}d\alpha e^{\alpha\frac{ix_{q}\left(  \vec{z}_{21} \cdot \vec{p}_{g}\right)
}{x_{_{\bar{q}}}+x_{q}}}%
\nonumber \\
&&\hspace{-.3cm}\times\left(  i(\vec{p}_{g} \cdot \vec{\varepsilon}_{g}^{\,\,\ast})\frac{Z_{q\bar
{q}g}K_{1}\left(  Q_{g}\left(  \alpha\right)  Z_{q\bar{q}g}\right)  }%
{Q_{g}\left(  \alpha\right)  }+x_{g}x_{q}(\vec{z}_{21} \cdot \vec{\varepsilon}%
_{g}^{\,\,\ast})K_{0}\left(  Q_{g}\left(  \alpha\right)  Z_{q\bar{q}g}\right)
\right)  -\left(  q\leftrightarrow\bar{q}\right)  \,,
\eqa
and
\beqa
\label{intF1L2}
&&\hspace{-.4cm} \int d\vec{z}_{3} \, F_{1}\left(  p_{q},p_{\bar{q}},p_{g},k,\vec{z}_{3},\vec
{z}_{2},\vec{z}_{1}\right)  ^{\alpha}\varepsilon_{L\alpha}=\delta_{\lambda
_{q},-\lambda_{\bar{q}}}\delta(k^{+}-p_{g}^{+}-p_{q}^{+}-p_{_{\bar{q}}}%
^{+})\theta(p_{g}^{+}-\sigma)\frac{2Qg}{\sqrt{2p_{g}^{+}}}%
\nonumber\\
&&\hspace{-.4cm}\times\frac{e^{-i\vec{p}_{_{\bar{q}}} \cdot \vec{z}_{_{2}}-i\vec{p}_{g} \cdot \vec{z}_{1}}%
}{(1-x_{q})x_{q}}\left[  (x_{_{\bar{q}}}+x_{g}\delta_{-s_{g}\lambda_{q}}%
)x_{q}\frac{2(\vec{z}_{12} \cdot \vec{\varepsilon}_{g}^{\,\,\ast})}{\vec{z}%
_{12}^{\,\,2}}e^{-i\frac{x_{_{\bar{q}}}(\vec{p}_{q} \cdot \vec{z}_{2})+x_{g}(\vec
{p}_{q} \cdot \vec{z}_{1})}{x_{g}+x_{_{\bar{q}}}}}\frac{Z_{\bar{q}g}K_{1}\left(
Z_{\bar{q}g}Q_{q}\left(  1\right)  \right)  }{Q_{q}\left(  1\right)  }\right.
\nonumber \\
&&\hspace{-.4cm}-(x_{q}+x_{g}\delta_{-s_{g}\lambda_{\bar{q}}})x_{_{\bar{q}}}e^{-i\vec{p}%
_{q} \cdot \vec{z}_{1}}%
\nonumber \\
&&\hspace{-.4cm}\left.  \times\!\int_{0}^{1}\!\!d\alpha \, e^{\alpha\frac{ix_{\bar{q}}\left(  \vec
{z}_{12} \cdot \vec{p}_{q}\right)  }{x_{_{g}}+x_{\bar{q}}}}\!\!\left(  i(\vec{p}_{q}%
 \cdot \vec{\varepsilon}_{g}^{\,\,\ast})\frac{Z_{\bar{q}gq}\left(  \alpha\right)
}{Q_{q}\left(  \alpha\right)  }K_{1}\left(  Q_{q}\left(  \alpha\right)
Z_{\bar{q}gq}\right)  -x_{q}x_{\bar{q}}(\vec{z}_{21} \cdot \vec{\varepsilon}%
_{g}^{\,\,\ast})K_{0}\left(  Q_{q}\left(  \alpha\right)  Z_{\bar{q}gq}\right)\!
\right) \! \right]  . \nonumber \\
\eqa
Note that
\begin{equation}
\int d\vec{z}_{3} \, F_{1}\left(  p_{q},p_{\bar{q}},p_{g},k,\vec{z}_{1},\vec
{z}_{3},\vec{z}_{2}\right)  ^{\alpha}\varepsilon_{L\alpha}=-\int d\vec{z}%
_{3}F_{1}\left(  p_{q},p_{\bar{q}},p_{g},k,\vec{z}_{3},\vec{z}_{2},\vec{z}%
_{1}\right)  ^{\alpha}\varepsilon_{L\alpha}|_{q\leftrightarrow\bar{q}}.
\end{equation}
In eqs.~(\ref{intF1L1},\ref{intF1L2}) we use the notations
\begin{equation}
Z_{ij}=\sqrt{\frac{x_{i}x_{_{j}}}{x_{_{i}}+x_{_{j}}}\vec{z}_{12}^{\,\,2}%
},\quad Z_{ijk}=\sqrt{x_{i}\frac{x_{j}+(1-\alpha)x_{i}x_{k}}{x_{j}+x_{i}}%
\vec{z}_{21}^{\,\,2}},\quad Q_{i}\left(  \alpha\right)  =\sqrt{\frac
{\alpha\vec{p}_{i}^{\,\,2}}{(1-x_{i})x_{i}}+Q^{2}}.
\end{equation}
For forward production $\vec{p}_{q}=\vec{p}_{\bar{q}}=\vec{p}_{g}=0$ one can
simplify these expressions with the help of eq.~(\ref{int_zK0_f})%
\beqa
&&\hspace{-1.2cm}\tilde{F}_{1}\left(  p_{q},p_{\bar{q}},p_{g},k,\vec{z}_{1},\vec{z}_{2}\right)
^{\alpha}\varepsilon_{L\alpha}|_{\vec{p}_{q}=\vec{p}_{\bar{q}}=\vec{p}_{g}%
=0}=-\delta(k^{+}-p_{g}^{+}-p_{q}^{+}-p_{_{\bar{q}}}^{+})\theta(p_{g}%
^{+}-\sigma)
\nonumber \\
&&\hspace{-1.2cm}\times\delta_{\lambda_{q},-\lambda_{\bar{q}}}\frac{4gN_{c}^{2}}{\sqrt
{2p_{g}^{+}}}\frac{(\vec{z}_{21}\cdot\vec{\varepsilon}_{g}^{\,\,\ast})}{\vec
{z}_{21}^{\,\,2}}\left\{  -\frac{Z_{q\bar{q}}K_{1}\left(  QZ_{q\bar{q}%
}\right)  }{2N_{c}^{2}x_{g}}+\left(  x_{_{\bar{q}}}+x_{g}\delta_{-s_{g}%
\lambda_{q}}\right)  \right.
\nonumber \\
&&\hspace{-1.2cm}\left.  \times\left(  \frac{Z_{g\bar{q}}K_{1}\left(  Z_{\bar{q}g}Q\right)
}{(1-x_{q})}-\frac{Z_{qg}K_{1}\left(  QZ_{qg}\right)  }{x_{\bar{q}}}%
+Z_{122}K_{1}\left(  QZ_{122}\right)  \left(  \frac{1}{x_{\bar{q}}}+\frac
{1}{N_{c}^{2}x_{g}}\right)  \right)  \right\}  -\left(  q\leftrightarrow
\bar{q}\right)  .
\label{resF1LtildeForward}
\eqa
For transverse photon (\ref{eT}) we can rewrite (\ref{F1eT}) as%
\beqa
&&\hspace{-1cm}F_{1}\left(  p_{q},p_{\bar{q}},p_{g},k,\vec{z}_{1},\vec{z}_{2},\vec{z}%
_{3}\right)  ^{\alpha}\varepsilon_{T\alpha} \nonumber \\
&&\hspace{-1cm}=-2ig\delta(k^{+}-p_{g}^{+}%
-p_{q}^{+}-p_{_{\bar{q}}}^{+})\theta(p_{g}^{+}-\sigma)\frac{e^{-i\vec{p}%
_{q} \cdot \vec{z}_{1}-i\vec{p}_{_{\bar{q}}} \cdot \vec{z}_{_{2}}-i\vec{p}_{g} \cdot \vec{z}_{3}}%
}{\pi\sqrt{2p_{g}^{+}}}\varepsilon_{g}^{\ast j}\varepsilon_{T}^{u}
\delta_{\lambda_{q},-\lambda_{\bar{q}}}\frac{z_{23}{}^{j}}{z_{23}{}^{2}%
}
\nonumber \\
&&\hspace{-1cm}\times\left(  \nabla_{1}^{u}\left(  \delta_{s\text{$\lambda_{\bar{q}}$}}\left(
x_{\bar{q}}+x_{g}\delta_{-s_{g}\lambda_{q}}\right)  -x_{q}\delta_{s\lambda
_{q}}\right)  -x_{q}\nabla_{3}^{u}\delta_{s\lambda_{q}}\delta_{-s_{g}%
\text{$\lambda_{\bar{q}}$}}\right)  K_{0}(QZ_{123})-\left(  q\leftrightarrow
\bar{q}\right)  .
\eqa
Then using (\ref{int_zK0}) and (\ref{int_zdK0}) we have%
\beqa
\label{intF1T1}
&&\int d\vec{z}_{3}F_{1}\left(  p_{q},p_{\bar{q}},p_{g},k,\vec{z}_{1},\vec
{z}_{2},\vec{z}_{3}\right)  ^{\alpha}\varepsilon_{T\alpha}=2ig\delta
(k^{+}-p_{g}^{+}-p_{q}^{+}-p_{_{\bar{q}}}^{+})\theta(p_{g}^{+}-\sigma
)\varepsilon_{g}^{\ast j}\varepsilon_{T}^{u}%
\nonumber \\
&&\times\delta_{\lambda_{q},-\lambda_{\bar{q}}}\frac{e^{-i\vec{p}_{q} \cdot \vec{z}%
_{1}-i\vec{p}_{_{\bar{q}}} \cdot \vec{z}_{_{2}}}}{\sqrt{2p_{g}^{+}}}\int_{0}%
^{1}d\alpha\left[  e^{-\frac{i\alpha x_{q}\left(  \vec{p}_{g} \cdot \vec{z}%
_{12}\right)  }{x_{\bar{q}}+x_{q}}-i\vec{p}_{g} \cdot \vec{z}_{2}}\left\{  \left(
\delta_{s\text{$\lambda_{\bar{q}}$}}\left(  x_{\bar{q}}+x_{g}\delta
_{-s_{g}\lambda_{q}}\right)  -x_{q}\delta_{s\lambda_{q}}\right)  \frac{{}}{{}%
}\right.  \right.
\nonumber \\
&&\times\left(  \frac{K_{0}\left(  Q_{g}\left(  \alpha\right)  Z_{q\bar{q}%
g}\right)  }{x_{\bar{q}}+x_{q}}\left(  \frac{ip_{g}{}^{j}z_{12}{}^{u}%
Z_{q\bar{q}g}{}^{2}}{\vec{z}_{12}{}^{2}x_{g}}-\frac{\alpha x_{q}{}^{2}ip_{g}%
{}^{u}z_{12}{}^{j}}{\left(  x_{\bar{q}}+x_{q}\right)  {}}+x_{q}\delta
^{ju}\right)  \right.
\nonumber \\
&&
\left.  -\frac{x_{q}Q_{g}\left(  \alpha\right)  Z_{q\bar{q}g}K_{1}\left(
Q_{g}\left(  \alpha\right)  Z_{q\bar{q}g}\right)  }{x_{\bar{q}}+x_{q}}\left(
\frac{\alpha p_{g}{}^{j}p_{g}{}^{u}}{Q_{g}^{2}\left(  \alpha\right)
x_{g}\left(  x_{\bar{q}}+x_{q}\right)  {}}+\frac{z_{12}{}^{j}z_{12}{}^{u}%
}{\vec{z}_{12}{}^{2}}\right)  \right)  -x_{q}\delta_{s\lambda_{q}}%
\delta_{-s_{g}\text{$\lambda_{\bar{q}}$}}%
\nonumber \\
&&\times\left(  \frac{Z_{q\bar{q}g}Q_{g}\left(  \alpha\right)  K_{1}\left(
Q_{g}\left(  \alpha\right)  Z_{q\bar{q}g}\right)  }{\left(  x_{\bar{q}}%
+x_{q}\right)  }\left(  \frac{\alpha p_{g}{}^{j}p_{g}{}^{u}}{Q_{g}^{2}\left(
\alpha\right)  x_{g}}+\frac{z_{12}{}^{j}z_{12}{}^{u}}{Z_{q\bar{q}g}^{2}%
}(1-\alpha)x_{g}x_{q}^{2}\right)  \right.
\nonumber \\
&&\left.  \left.  \left.  +K_{0}\left(  Q_{g}\left(  \alpha\right)  Z_{q\bar
{q}g}\right)  \left(  \frac{\alpha x_{q}z_{12}{}^{j}ip_{g}{}^{u}}{x_{\bar{q}%
}+x_{q}}+z_{12}{}^{u}ip_{g}{}^{j}\frac{(\alpha-1)x_{q}}{x_{\bar{q}}+x_{q}%
}-\delta^{ju}\right)  \right)  \right\}  -\left(  q\leftrightarrow\bar
{q}\right)  \right]  \,,
\eqa
which can be also put in the form
\beqa
\label{intF1T2}
&&\int d\vec{z}_{3}F_{1}\left(  p_{q},p_{\bar{q}},p_{g},k,\vec{z}_{3},\vec
{z}_{2},\vec{z}_{1}\right)  ^{\alpha}\varepsilon_{T\alpha} \nonumber \\
&&=2ig\delta
(k^{+}-p_{g}^{+}-p_{q}^{+}-p_{_{\bar{q}}}^{+})\theta(p_{g}^{+}-\sigma
)\varepsilon_{g}^{\ast j}\varepsilon_{T}^{u}\delta_{\lambda_{q},-\lambda
_{\bar{q}}} \frac{e^{-i\vec{p}_{g} \cdot \vec{z}_{1}-i\vec{p}_{_{\bar{q}}} \cdot \vec{z}_{_{2}}}%
}{\sqrt{2p_{g}^{+}}} %
\nonumber \\
&&
\times\left[  2e^{-i\frac{x_{\bar{q}}\left(  \vec{p}_{q} \cdot \vec
{z}_{2}\right)  +x_{g}\left(  \vec{p}_{q} \cdot \vec{z}_{1}\right)  }{x_{\bar{q}%
}+x_{g}}}\left\{  \left(  \delta_{s\text{$\lambda_{\bar{q}}$}}\left(
x_{\bar{q}}+x_{g}\delta_{-s_{g}\lambda_{q}}\right)  -x_{q}\delta_{s\lambda
_{q}}\right)  \frac{z_{12}{}^{j}ip_{q}{}^{u}Z_{\bar{q}g}K_{1}\left(
Z_{\bar{q}g}Q_{q}(1)\right)  }{z_{12}{}^{2}Q_{q}(1)x_{q}\left(  x_{\bar{q}%
}+x_{g}\right)  }\right.  \right.
\nonumber \\
&&\left.  +\frac{\delta_{s\lambda_{q}}\delta_{-s_{g}\text{$\lambda_{\bar{q}}$}%
}x_{g}}{\left(  x_{\bar{q}}+x_{g}\right)  {}^{2}}\frac{z_{12}{}^{j}}{z_{12}%
{}^{2}}\left(  z_{12}{}^{u}x_{\bar{q}}K_{0}\left(  Z_{\bar{q}g}Q_{q}%
(1)\right)  +\frac{ip_{q}{}^{u}}{Q_{q}(1)}Z_{\bar{q}g}K_{1}\left(  Z_{\bar
{q}g}Q_{q}(1)\right)  \right)  \right\}
\nonumber \\
&&-\int_{0}^{1}d\alpha e^{i\frac{\alpha x_{\bar{q}}\left(  \vec{p}_{q} \cdot \vec
{z}_{12}\right)  }{x_{\bar{q}}+x_{g}}-i\vec{p}_{q} \cdot \vec{z}_{1}}\left\{  \left(
x_{\bar{q}}\delta_{s\text{$\lambda_{\bar{q}}$}}\delta_{-s_{g}\lambda_{q}%
}+\left(  \delta_{s\lambda_{q}}\left(  x_{g}\delta_{-s_{g}\text{$\lambda
_{\bar{q}}$}}+x_{q}\right)  -x_{\bar{q}}\delta_{s\text{$\lambda_{\bar{q}}$}%
}\right)  \right)  \frac{{}}{{}}\right.
\nonumber \\
&&
\times\left(  \frac{x_{\bar{q}}Q_{q}\left(  \alpha\right)  Z_{\bar{q}gq}%
K_{1}\left(  Q_{q}\left(  \alpha\right)  Z_{\bar{q}gq}\right)  }{x_{\bar{q}%
}+x_{g}}\left(  \frac{\alpha p_{q}{}^{j}p_{q}{}^{u}}{Q_{q}^{2}\left(
\alpha\right)  x_{q}\left(  x_{\bar{q}}+x_{g}\right)  {}}+\frac{z_{12}{}%
^{j}z_{12}{}^{u}}{z_{12}{}^{2}}\right)  \right.
\nonumber \\
&&\left.  +\frac{K_{0}\left(  Q_{q}\left(  \alpha\right)  Z_{\bar{q}gq}\right)
}{x_{\bar{q}}+x_{g}}\left(  \frac{ip_{q}{}^{j}z_{12}{}^{u}Z_{\bar{q}gq}{}^{2}%
}{x_{q}z_{12}{}^{2}}-\frac{z_{12}{}^{j}ip_{q}{}^{u}\alpha x_{\bar{q}}{}^{2}%
}{x_{\bar{q}}+x_{g}{}}-x_{\bar{q}}\delta^{ju}\right)  \right)  +x_{\bar{q}%
}\delta_{s\lambda_{\bar{q}}}\delta_{-s_{g}\lambda_{q}}%
\nonumber \\
&&\times\left(  -\frac{Q_{q}\left(  \alpha\right)  Z_{\bar{q}gq}K_{1}\left(
Q_{q}\left(  \alpha\right)  Z_{\bar{q}gq}\right)  }{x_{\bar{q}}+x_{g}}\left(
\frac{\alpha p_{q}{}^{j}p_{q}{}^{u}}{Q_{q}^{2}\left(  \alpha\right)  x_{q}%
}+\frac{z_{12}{}^{j}z_{12}{}^{u}}{Z_{\bar{q}gq}^{2}}(1-\alpha)x_{q}x_{\bar{q}%
}^{2}\right)  \right.
\nonumber \\
&&\left.  \left.  \left.  +K_{0}\left(  Q_{q}\left(  \alpha\right)  Z_{\bar
{q}gq}\right)  \left(  \frac{ip_{q}{}^{u}z_{12}{}^{j}\alpha x_{\bar{q}}%
}{x_{\bar{q}}+x_{g}}+\delta^{ju}-ip_{q}{}^{j}z_{12}{}^{u}\frac{x_{\bar{q}}%
}{x_{\bar{q}}+x_{g}}(1-\alpha)\right)  \right)  \right\}  \right]  \,.
\eqa
Note that
\begin{equation}
\int d\vec{z}_{3}F_{1}\left(  p_{q},p_{\bar{q}},p_{g},k,\vec{z}_{1},\vec
{z}_{3},\vec{z}_{2}\right)  ^{\alpha}\varepsilon_{T\alpha}=-\int d\vec{z}%
_{3}F_{1}\left(  p_{q},p_{\bar{q}},p_{g},k,\vec{z}_{3},\vec{z}_{2},\vec{z}%
_{1}\right)  ^{\alpha}\varepsilon_{T\alpha}|_{q\leftrightarrow\bar{q}}\,.
\end{equation}
For forward production $\vec{p}_{q}=\vec{p}_{\bar{q}}=\vec{p}_{g}=0$ one can
simplify these expressions via eqs.~(\ref{int_zK0_f}) and (\ref{int_zdK0_f}), thus obtaining
\beqa
\label{resF1TtildeForward}
&&\hspace{-1cm}\int d\vec{z}_{3}F_{1}\left(  p_{q},p_{\bar{q}},p_{g},k,\vec{z}_{1},\vec
{z}_{2},\vec{z}_{3}\right)  ^{\alpha}\varepsilon_{T\alpha}=2ig\delta
(k^{+}-p_{g}^{+}-p_{q}^{+}-p_{_{\bar{q}}}^{+})\theta(p_{g}^{+}-\sigma
)\frac{\delta_{-\lambda_{\bar{q}}\lambda_{q}}}{x_{g}\sqrt{2p_{g}^{+}}}%
\nonumber \\
&&\hspace{-1cm}\times\left[  \delta_{-s_{g}s}\frac{(\vec{z}_{12} \cdot \vec{\varepsilon}%
_{g}^{\,\,\ast})(\vec{\varepsilon}_{T} \cdot \vec{z}_{12}{})}{\vec{z}_{12}{}^{2}%
}\left(  K_{2}\left(  QZ_{q\bar{q}}\right)  +2\left(  x_{q}-1\right)
K_{2}\left(  QZ_{122}\right)  \right)  \left(  x_{q}\delta_{s\lambda_{q}%
}-x_{\bar{q}}\delta_{s\text{$\lambda_{\bar{q}}$}}\right)  \right.
\nonumber \\
&&\hspace{-1cm}\left.  -\delta_{ss_{g}}\left(  K_{0}\left(  QZ_{122}\right)  \left(
x_{q}x_{\bar{q}}\delta_{s\lambda_{q}}-\left(  x_{q}-1\right)  {}^{2}%
\delta_{s\text{$\lambda$}_{\bar{q}}}\right)  +\frac{x_{\bar{q}}\delta
_{s\text{$\lambda$}_{\bar{q}}}}{x_{\bar{q}}+x_{q}}K_{0}\left(  QZ_{q\bar{q}%
}\right)  \right)  \right]  -\left(  q\leftrightarrow\bar{q}\right)  \,,
\eqa
which can be also put in the form
\beqa
\label{resF1TtildeForward2}
&&\hspace{-.2cm}\int d\vec{z}_{3}F_{1}\left(  p_{q},p_{\bar{q}},p_{g},k,\vec{z}_{3},\vec
{z}_{2},\vec{z}_{1}\right)  ^{\alpha}\varepsilon_{T\alpha}=2ig\delta
(k^{+}-p_{g}^{+}-p_{q}^{+}-p_{_{\bar{q}}}^{+})\theta(p_{g}^{+}-\sigma
)\frac{\delta_{-\lambda_{\bar{q}}\lambda_{q}}}{x_{q}\sqrt{2p_{g}^{+}}}%
\nonumber \\
&&\hspace{-.2cm}\times\left[  2\delta_{-s_{g}s}\frac{(\vec{z}_{12}{} \cdot \vec{\varepsilon}%
_{g}^{\,\,\ast})(\vec{\varepsilon}_{T} \cdot \vec{z}_{12}{})}{\vec{z}_{12}{}^{2}%
}\left(  \frac{x_{g}}{x_{q}-1}K_{2}\left(  QZ_{\bar{q}g}\right)  -\left(
x_{\bar{q}}-1\right)  K_{2}\left(  QZ_{121}\right)  \right)  \left(
x_{q}\delta_{s\lambda_{q}}-x_{\bar{q}}\delta_{s\text{$\lambda_{\bar{q}}$}%
}\right)  \right.\nonumber \\
&&\hspace{-.2cm}
\left.  -\delta_{ss_{g}}\left(  \frac{\delta_{s\lambda_{q}}x_{g}{}^{2}%
}{\left(  x_{q}-1\right)  {}^{2}}K_{0}\left(  QZ_{\bar{q}g}\right)  +\left(
x_{q}x_{\bar{q}}\delta_{s\text{$\lambda_{\bar{q}}$}}-\left(  x_{\bar{q}%
}-1\right)  {}^{2}\delta_{s\lambda_{q}}\right)  K_{0}\left(  QZ_{121}\right)
\right)  \right]  . 
\eqa
The expressions (\ref{resF1LtildeForward}) and (\ref{resF1TtildeForward2}) can be used as a starting point for the description of e.g. meson production within the QCD collinear
factorization, see ref.~\cite{Besse:2012ia}.

\section{Impact factor in the momentum space and in the linear approximation}
\label{Sec:momentum}

Here we will calculate the Fourier transform of the impact factors. We can
rewrite the matrix element (\ref{M}) as
\beqa
&&\hspace{-.3cm}M^{\alpha} \nonumber \\
&&\hspace{-.3cm}=\!\!\int \! d\vec{p}_{1}d\vec{p}_{2}d\vec{p}_{3}\,F_{1}\left(  p_{q}%
,p_{\bar{q}},p_{g},z_{0},\vec{p}_{1},\vec{p}_{2},\vec{p}_{3}\right)  ^{\alpha
}\frac{1}{2}\left[  N_{c}^{2}\left(  \mathbf{U}_{32}+\mathbf{U}_{13}%
+\mathbf{U}_{32}\mathbf{U}_{13}\right)  -\mathbf{U}_{12}\right]  \left(
\vec{p}_{1},\vec{p}_{2},\vec{p}_{3}\right)
\nonumber \\
&&\hspace{-.3cm}
+\int d\vec{p}_{1}d\vec{p}_{2}\,\tilde{F}_{2}\left(  p_{q},p_{\bar{q}}%
,p_{g},z_{0},\vec{p}_{1},\vec{p}_{2}\right)  ^{\alpha}\frac{N_{c}^{2}-1}%
{2}\mathbf{U}\left(  \vec{p}_{1},\vec{p}_{2}\right)  .
\eqa
Here the Fourier transforms are defined as%
\begin{equation}
F_{1}\left(  p_{q},p_{\bar{q}},p_{g},z_{0},\vec{p}_{1},\vec{p}_{2},\vec{p}%
_{3}\right)  ^{\alpha}=\int\frac{d\vec{z}_{1}}{2\pi}\frac{d\vec{z}_{2}}{2\pi
}\frac{d\vec{z}_{3}}{2\pi} \, e^{i[\vec{p}_{1} \cdot \vec{z}_{1}+\vec{p}_{2} \cdot \vec{z}%
_{2}+\vec{p}_{3} \cdot \vec{z}_{3}]}F_{1}\left(  p_{q},p_{\bar{q}},p_{g},z_{0}%
,\vec{z}_{1},\vec{z}_{2},\vec{z}_{3}\right)  ^{\alpha},
\end{equation}%
\begin{equation}
\tilde{F}_{2}\left(  p_{q},p_{\bar{q}},p_{g},z_{0},\vec{p}_{1},\vec{p}%
_{2}\right)  ^{\alpha}=\int\frac{d\vec{z}_{1}}{2\pi}\frac{d\vec{z}_{2}}{2\pi
}e^{i[\vec{p}_{1} \cdot \vec{z}_{1}+\vec{p}_{2} \cdot \vec{z}_{2}]}\tilde{F}_{2}\left(
p_{q},p_{\bar{q}},p_{g},z_{0},\vec{z}_{1},\vec{z}_{2}\right)  ^{\alpha},
\end{equation}
and%
\[
\left[  N_{c}^{2}\left(  \mathbf{U}_{32}+\mathbf{U}_{13}+\mathbf{U}%
_{32}\mathbf{U}_{13}\right)  -\mathbf{U}_{12}\right]  \left(  \vec{p}_{1}%
,\vec{p}_{2},\vec{p}_{3}\right)  =\int\frac{d\vec{z}_{1}}{2\pi}\frac{d\vec
{z}_{2}}{2\pi}\frac{d\vec{z}_{3}}{2\pi}e^{-i[\vec{p}_{1} \cdot \vec{z}_{1}+\vec
{p}_{2} \cdot \vec{z}_{2}+\vec{p}_{3} \cdot \vec{z}_{3}]}%
\]%
\begin{equation}
\times\left[  N_{c}^{2}\left(  \mathbf{U}_{32}+\mathbf{U}_{13}+\mathbf{U}%
_{32}\mathbf{U}_{13}\right)  -\mathbf{U}_{12}\right]  ,
\end{equation}%
\begin{equation}
\mathbf{U}\left(  \vec{p}_{1},\vec{p}_{2}\right)  =\int\frac{d\vec{z}_{1}%
}{2\pi}\frac{d\vec{z}_{2}}{2\pi}e^{-i[\vec{p}_{1} \cdot \vec{z}_{1}+\vec{p}_{2}%
 \cdot \vec{z}_{2}]}\mathbf{U}_{12}.
\end{equation}
To get the linearized impact factor, one should neglect the term $\mathbf{U}%
_{32}\mathbf{U}_{13}$ and write%
\beqa
\label{linU}
&&\left[  N_{c}^{2}\left(  \mathbf{U}_{32}+\mathbf{U}_{13}+\mathbf{U}%
_{32}\mathbf{U}_{13}\right)  -\mathbf{U}_{12}\right]  \left(  \vec{p}_{1}%
,\vec{p}_{2},\vec{p}_{3}\right)
\nonumber \\
&&
\sim 2\pi\left[  N_{c}^{2}\left(  \delta(\vec{p}_{1})\mathbf{U}\left(
\vec{p}_{3},\vec{p}_{2}\right)  +\delta(\vec{p}_{2})\mathbf{U}\left(  \vec
{p}_{1},\vec{p}_{3}\right)  \right)  -\delta(\vec{p}_{3})\mathbf{U}\left(
\vec{p}_{1},\vec{p}_{2}\right)  \right]  .
\eqa
Then for the matrix element $M^{\alpha}$ we get%
\begin{equation}
M^{\alpha}\!=\frac{1}{2}\!\int \! d\vec{p}_{1}d\vec{p}_{2} \, \mathbf{U}\!\!\left(  \vec
{p}_{1},\vec{p}_{2}\right)  \left\{  \tilde{F}_{1}\left(  p_{q},p_{\bar{q}%
},p_{g},z_{0},\vec{p}_{1},\vec{p}_{2}\right)  ^{\alpha}\!+\!\left(  N_{c}%
^{2}-1\right)  \tilde{F}_{2}\left(  p_{q},p_{\bar{q}},p_{g},z_{0},\vec{p}%
_{1},\vec{p}_{2}\right)  ^{\alpha}\right\}  ,
\end{equation}%
\beq
\label{linF1tilde}
\tilde{F}_{1}\left(  p_{q},p_{\bar{q}},p_{g},z_{0},\vec{p}_{1},\vec{p}%
_{2}\right)  ^{\alpha}=2\pi N_{c}^{2}F_{1}\left(  p_{q},p_{\bar{q}}%
,p_{g},z_{0},0,\vec{p}_{2},\vec{p}_{1}\right)  ^{\alpha}%
\eq
\begin{equation}
+2\pi N_{c}^{2}F_{1}\left(  p_{q},p_{\bar{q}},p_{g},z_{0},\vec{p}_{1}%
,0,\vec{p}_{2}\right)  ^{\alpha}-2\pi F_{1}\left(  p_{q},p_{\bar{q}}%
,p_{g},z_{0},\vec{p}_{1},\vec{p}_{2},0\right)  ^{\alpha}.
\end{equation}
Taking the Fourier transform of eq.~(\ref{F1eL}) via eq.~(\ref{int_dqzK0_123}) from
appendix B we get for the longitudinal photon%
\beqa
\label{F1Llin}
&&F_{1}\left(  p_{q},p_{\bar{q}},p_{g},z_{0},\vec{p}_{1},\vec{p}_{2},\vec{p}%
_{3}\right)  ^{\alpha}\varepsilon_{L\alpha}=\delta(k^{+}-p_{g}^{+}-p_{q}%
^{+}-p_{_{\bar{q}}}^{+})\delta\left(  \vec{p}_{1q}+\vec{p}_{2\bar{q}}+\vec
{p}_{3g}\right)  \theta(p_{g}^{+}-\sigma)
\nonumber \\
&&\times\frac{\delta_{\lambda_{q},-\lambda_{\bar{q}}}}{\sqrt{2p_{g}^{+}}}%
\frac{4iQ\, g \,(x_{q}+x_{g}\delta_{-s_{g}\lambda_{\bar{q}}})\left(  (\vec{p}%
_{2\bar{q}} \cdot \vec{\varepsilon}_{g}^{\,\,\ast}){}x_{q}+(\vec{p}_{1q}%
 \cdot \vec{\varepsilon}_{g}^{\,\,\ast}){}\left(  1-x_{\bar{q}}\right)  \right)
}{\left(  1-x_{\bar{q}}\right)  x_{g}x_{q}\left(  Q^{2}+\frac{\vec{p}%
_{2\bar{q}}{}^{2}}{x_{\bar{q}}(1-x_{\bar{q}}{})}\right)  \left(  Q^{2}%
+\frac{\vec{p}_{1q}{}^{2}}{x_{q}}+\frac{\vec{p}_{2\bar{q}}{}^{2}}{x_{\bar{q}}%
}+\frac{\vec{p}_{3g}{}^{2}}{x_{g}}\right)  }-(q\leftrightarrow\bar{q}),
\eqa
where%
\begin{equation}
\left(  q\leftrightarrow\bar{q}\right)  \equiv\left(  \lambda_{q},x_{q}%
,\vec{p}_{1},\vec{p}_{q}\leftrightarrow\lambda_{\bar{q}},x_{\bar{q}},\vec
{p}_{2},\vec{p}_{\bar{q}}\right)  .
\end{equation}
One can check that this result is compatible with the wave function derived in ref.~\cite{Beuf:2011xd}.
Using eq.~(\ref{int_dqzK0_122}), we get%
\beqa
\label{F2tildeLlin}
&&\tilde{F}_{2}\left(  p_{q},p_{\bar{q}},p_{g},k,\vec{p}_{1},\vec{p}_{2}\right)
^{\alpha}\varepsilon_{L\alpha}=\frac{4igQ}{\sqrt{2p_{g}^{+}}}\theta(p_{g}%
^{+}-\sigma)\delta(k^{+}-p_{g}^{+}-p_{q}^{+}-p_{_{\bar{q}}}^{+})\delta(\vec
{p}_{1q}+\vec{p}_{2\bar{q}}-\vec{p}_{g})
\nonumber \\
&&\times\delta_{\lambda_{q},-\lambda_{\bar{q}}}\frac{x_{q}\left(  x_{g}%
+x_{\bar{q}}\right)  \left(  \delta_{-s_{g}\lambda_{q}}x_{g}+x_{\bar{q}%
}\right)  }{\left(  p_{g}+p_{\bar{q}}\right)  {}^{2}}\frac{2\pi(\vec{P}%
_{\bar{q}} \cdot \vec{\varepsilon}_{g}^{\,\,\ast})}{(Q^{2}x_{q}(x_{_{\bar{q}}%
}+x_{g})+\vec{p}_{1q}^{\,\,2})}-\left(  q\leftrightarrow\bar{q}\right)  .
\eqa
For the transverse photon, using eqs.~(\ref{int_dqzz1z2K1_123}) and
(\ref{int_dqzz2z2K1_123}) we have%
\beqa
\label{F1Tlin}
&&\hspace{-.3cm}F_{1}\left(  p_{q},p_{\bar{q}},p_{g},z_{0},\vec{p}_{1},\vec{p}_{2},\vec{p}%
_{3}\right)  ^{\alpha}\varepsilon_{T\alpha}=\frac{-4ig}{\sqrt{2p_{g}^{+}}%
}\delta(k^{+}\!-p_{g}^{+}\!-p_{q}^{+}\!-p_{_{\bar{q}}}^{+})\delta\left(  \vec
{p}_{1q}+\vec{p}_{2\bar{q}}+\vec{p}_{3g}\right)  \theta(p_{g}^{+}-\sigma)
\nonumber \\
&&\hspace{-.3cm}\times\delta_{\lambda_{q},-\lambda_{\bar{q}}}\frac{(\vec{p}_{2\bar{q}}{}%
 \cdot \vec{\varepsilon}_{g}^{\,\,\ast})\left(  x_{\bar{q}}+x_{g}\right)  +(\vec
{p}_{1q} \cdot \vec{\varepsilon}_{g}^{\,\,\ast})\,x_{\bar{q}}}{(\vec{p}_{2\bar{q}}%
{}\left(  x_{\bar{q}}+x_{g}\right)  +\vec{p}_{1q}{}x_{\bar{q}})^{2}}\left\{
\frac{\left(  x_{q}-\delta_{s\lambda_{\bar{q}}}\right)  \left(  x_{\bar{q}%
}+x_{g}\delta_{-s_{g}\lambda_{q}}\right)  }{x_{q}}\right.
\nonumber \\
&&\hspace{-.3cm}\times\left(  \frac{(\vec{p}_{1q}{} \cdot \vec{\varepsilon}_{T})\left(  x_{g}%
+x_{q}\right)  +(\vec{p}_{2\bar{q}}{} \cdot \vec{\varepsilon}_{T}) \, x_{q}}{x_{g}\left(
\frac{\vec{p}_{2\bar{q}}{}^{2}}{x_{\bar{q}}}+\frac{\vec{p}_{3g}{}^{2}}{x_{g}%
}+\frac{\vec{p}_{1q}{}^{2}}{x_{q}}+Q^{2}\right)  }-\frac{(\vec{p}_{1q} \cdot 
\vec{\varepsilon}_{T})}{\left(  x_{\bar{q}}+x_{g}\right)  \left(  \frac
{\vec{p}_{1q}{}^{2}}{x_{q}\left(  x_{\bar{q}}+x_{g}\right)  }+Q^{2}\right)
}\right)
\nonumber \\
&&\hspace{-.3cm}\left.  +\left(  x_{\bar{q}}+x_{g}\delta_{-s_{g}\lambda_{q}}-\delta
_{s\lambda_{q}}\right)  \frac{(\vec{p}_{2\bar{q}}{} \cdot \vec{\varepsilon}%
_{T})\left(  x_{\bar{q}}+x_{g}\right)  +(\vec{p}_{1q}{} \cdot \vec{\varepsilon}%
_{T}) \, x_{\bar{q}}}{x_{g}\left(  \frac{\vec{p}_{2\bar{q}}{}^{2}}{x_{\bar{q}}%
}+\frac{\vec{p}_{3g}{}^{2}}{x_{g}}+\frac{\vec{p}_{1q}{}^{2}}{x_{q}}%
+Q^{2}\right)  }\right\}  -\left(  q\leftrightarrow\bar{q}\right)
\nonumber \\
&&\hspace{-.3cm}=\frac{2ig}{\sqrt{2p_{g}^{+}}}\frac{\delta(k^{+}-p_{g}^{+}-p_{q}^{+}%
-p_{_{\bar{q}}}^{+})\delta\left(  \vec{p}_{1q}+\vec{p}_{2\bar{q}}+\vec{p}%
_{3g}\right)  \theta(p_{g}^{+}-\sigma)\delta_{-\lambda_{\bar{q}}\lambda_{q}}%
}{Q^{2}\left(  1-x_{q}\right)  \left(  \frac{\vec{p}_{2\bar{q}}{}^{2}}%
{x_{\bar{q}}}+\frac{\vec{p}_{3g}{}^{2}}{x_{g}}+\frac{\vec{p}_{1q}{}^{2}}%
{x_{q}}+Q^{2}\right)  }\left\{  \delta_{ss_{g}}\delta_{s\lambda_{q}}\frac{{}%
}{{}}\right.
\nonumber \\
&&\hspace{-.3cm}+\!\left.  2(\vec{p}_{1q} \cdot \vec{\varepsilon}_{T})((\vec{p}_{2\bar{q}}{}%
 \cdot \vec{\varepsilon}_{g}^{\,\,\ast})\!\left(  x_{\bar{q}}+x_{g}\right)  +(\vec
{p}_{1q}{} \cdot \vec{\varepsilon}_{g}^{\,\,\ast}) \,x_{\bar{q}})\frac{\left(
x_{q}-\delta_{s\lambda_{\bar{q}}}\right) \! \left(  x_{g}\delta_{-s_{g}%
\lambda_{q}}+x_{\bar{q}}\right)  }{\left(  1-x_{q}\right)  {}x_{q}x_{\bar{q}%
}x_{g}\!\left( \! Q^{2}+\frac{\vec{p}_{1q}{}^{2}}{\left(  1-x_{q}\right)  x_{q}%
}\!\right)  } \!\!\right\} \! -\left(  q\leftrightarrow\bar{q}\right) \! . \!\!\nonumber \\
\eqa
Again, one can check that this result is compatible with the wave function derived in ref.~\cite{Beuf:2011xd}.
Finally, using eq.~(\ref{int_dqzz1K1_122}) we find%
\beqa
\label{F2tildeTlin}
&&\hspace{-.25cm}\tilde{F}_{2}\left(  p_{q},p_{\bar{q}},p_{g},k,\vec{p}_{1},\vec{p}_{2}\right)
^{\alpha}\!\varepsilon_{T\alpha}=-4g\,\theta(p_{g}^{+}-\sigma)\,\delta(k^{+}%
\!-p_{g}^{+}\!-p_{q}^{+}\!-p_{_{\bar{q}}}^{+})\delta(\vec{p}_{1q}+\vec{p}_{2\bar{q}%
}-\vec{p}_{g})\frac{\delta_{\lambda_{q},-\lambda_{\bar{q}}}}{\sqrt{2p_{g}^{+}%
}}%
\nonumber \\
&&\times\frac{\left(  \delta_{\lambda_{\bar{q}}s}-x_{q}\right)  \left(
\delta_{-s_{g}\lambda_{q}}x_{g}+x_{\bar{q}}\right)  }{\left(  p_{g}+p_{\bar
{q}}\right)  {}^{2}}\frac{2\pi i(\vec{P}_{\bar{q}}\cdot\vec{\varepsilon}%
_{g}^{\,\,\ast})\left(  \vec{p}_{1q}\cdot \vec{\varepsilon}_{T}\right)  }{Q^{2}%
x_{q}(x_{_{\bar{q}}}+x_{g})+\vec{p}_{1q}^{\,\,2}}-\left(  q\leftrightarrow
\bar{q}\right)  .
\eqa
The formulas in the momentum space derived in this section constitute a convenient starting point for calculations of
phenomenologically important observables such as cross sections etc.

\section{Conclusions}
\label{Sec:Conclusion}

Based on the QCD shock-wave approach~\cite{Balitsky:1995ub,Balitsky:2010ze,Balitsky:2012bs}, we rederived 
the $\gamma^* \rightarrow q\bar{q}$ impact factor. Using the same approach, we computed 
 the general expression for the
$\gamma^*\rightarrow q\bar{q}g$ impact factor for the first time. The contribution of the diagrams with the gluons crossing the shock-wave, calculated using Balitsky's formalism, are consistent with the results for the
$\gamma^* \rightarrow q\bar{q}g$ wave function obtained in ref.~\cite{Beuf:2011xd}, based on old-fashioned perturbation theory.

The results we obtained, in coordinate space, are very suitable for phenomenological studies of diffractive processes since
they allow for the implementation of saturation models when considering the color-singlet channel. 
The measurement of dijet production in DDIS was recently performed~\cite{Aaron:2011mp}, and a precise comparison of 
dijet versus triple-jet production, which has not been  performed yet at HERA~\cite{Adloff:2000qi}, would be very useful to get a deeper understanding of the QCD mechanism underlying diffraction. Such a ratio would provide an observable 
possibly more independent of any saturation effect. A quantitative first principle analysis of this would  require an evaluation of virtual corrections to the $\gamma^* \rightarrow q\bar{q}$ impact factor, which are left for further studies.

Our results could also be relevant for photo-production of diffractive jets~\cite{Chekanov:2007rh,Aaron:2010su}, the hard scale being provided by the invariant mass of the produced state. Indeed, the direct coupling of a Pomeron to the impact factor could be important, in addition to the resolved Pomeron contribution (which is the sum of a direct interaction of the photon with
quarks or gluons originating from the pomeron, and a resolved
photon-pomeron interaction), in particular in the region $x_\gamma \sim 1$ ($x_\gamma$ is the longitudinal momentum fraction carried by the
partons coming from the photon), in view of the  collinear factorization breaking which has been the matter of discussions~\cite{Klasen:2004qr,Klasen:2008ah}. 
Since our results are expressed in terms of a shock-wave, they can 
be used both for inclusive (considering the color octet in the $t-$channel, by modifying formula (\ref{M}) and diffractive (in the color-singlet case) jet production, the ratio of amplitudes providing an interesting observable 
to evaluate gap survival probabilities~\cite{Kaidalov:2003xf,Klasen:2004qr,Klasen:2008ah}.

Furthermore,  our results, expressed in terms of a shock-wave,  are a natural starting point for studies of higher-twist effects, which could be investigated by an appropriate expansion of $U$ operators in powers of the coupling, in order to study the effect of multigluon exchange in the $t-$channel.

Finally, diffractive open charm production was measured at HERA~\cite{Aktas:2006up} 
and studied in the large $M$ limit based on the direct coupling between a Pomeron and a $q \bar{q}$ or a $q\bar{q}g$ state, with massive quarks~\cite{Bartels:2002ri}.
The extension of our result to the case of massive quark,  is  left for future analysis.

Our result is therefore
a first step for phenomenological studies of diffraction, which could be of relevance in future $e-p$ and $e-A$ colliders like EIC and LHeC, as well as for ultraperipheral processes which could be studied at LHC.

\section*{Acknowledgements}

We would like to thank A.~Besse, G.~Beuf, L.~Motyka, Al~Mueller, S.~Munier and M.~Sadzi-kowski for discussions.
A.~V.~G. thanks V.~S.~Fadin and A.~V.~Reznichenko for helpful discussions and the LPT Orsay and NCBJ in Warsaw for hospitality while part of
this work was being done. A.~V.~G. also acknowledges support of president
grant MK-525.2013.2 and RFBR grant 13-02-01023.
This work was partially supported by the PEPS-PTI PHENO-DIFF, the PRC0731 DIFF-QCD, the Polish Grant NCN
No. DEC-2011/01/B/ST2/03915 
and the Joint Research Activity Study of Strongly 
Interacting Matter (acronym HadronPhysics3, Grant Agreement n.283286) under the Seventh
Framework
Programme of the European Community.


\section*{Appendix A: Integrals necessary for linearization}
\label{Ap:A}

\qquad We need the following integrals%
\begin{equation}
\int d\vec{z}_{1}e^{-i\vec{p}_{q} \cdot \vec{z}_{1}}K_{0}\left(  QZ_{123}\right)
|_{\vec{z}_{3}\rightarrow\vec{z}_{1}}=\frac{2\pi}{x_{q}(1-x_{q})}%
e^{-i\frac{x_{_{\bar{q}}}(\vec{p}_{q} \cdot \vec{z}_{2})+x_{g}(\vec{p}_{q} \cdot \vec{z}%
_{1})}{x_{g}+x_{_{\bar{q}}}}}\frac{Z_{\bar{q}g}K_{1}\left(  Z_{\bar{q}g}%
Q_{q}\left(  1\right)  \right)  }{Q_{q}\left(  1\right)  }, \label{int_K0}%
\end{equation}
where we define
\begin{equation}
Z_{ij}=\sqrt{\frac{x_{i}x_{_{j}}}{x_{_{i}}+x_{_{j}}}\vec{z}_{12}^{\,\,2}%
},\quad Q_{i}\left(  \alpha\right)  =\sqrt{\frac{\alpha\vec{p}_{i}^{\,\,2}%
}{(1-x_{i})x_{i}}+Q^{2}} \,.
\end{equation}%
\beqa
\label{int_zK0}
&&\hspace{-.4cm}\int d\vec{z}_{3}e^{-i\vec{p}_{g}\cdot \vec{z}_{3}}\frac{\vec{z}_{32}}{\vec{z}%
_{32}^{\,\,2}}K_{0}\left(  QZ_{123}\right) %
 \\
&&\hspace{-.4cm} =-\pi ie^{-i\vec{p}_{g}\cdot\vec{z}%
_{2}}\int_{-\infty}^{0}dte^{\frac{itx_{g}x_{q}Q^{2}\vec{z}_{21}^{\,\,2}}%
{2}-i\frac{(x_{_{\bar{q}}}+x_{g})+i0}{2tx_{g}}}\frac{t\vec{p}_{g}+\frac
{\vec{z}_{21}}{\vec{z}_{21}^{\,\,2}}}{\left(  t\vec{p}_{g}+\frac{\vec{z}_{21}%
}{\vec{z}_{21}^{\,\,2}}\right)  ^{2}}\left(  e^{\frac{ix_{q}\vec{z}%
_{21}^{\,\,2}}{2(x_{_{\bar{q}}}+x_{q})t}\left(  t\vec{p}_{g}+\frac{\vec
{z}_{21}}{\vec{z}_{21}^{\,\,2}}\right)  ^{2}}-1\right)
\nonumber \\
&&\hspace{-.4cm}=-\frac{\pi e^{-i\vec{p}_{g}\cdot\vec{z}_{2}}}{(1-x_{g})x_{g}}\int_{0}^{1}d\alpha
e^{\alpha\frac{ix_{q}\left(  \vec{z}_{21}\cdot\vec{p}_{g}\right)  }{x_{_{\bar{q}}%
}+x_{q}}}\left(  \frac{i\vec{p}_{g}Z_{q\bar{q}g}}{Q_{g}\left(  \alpha\right)
}K_{1}\left(  Q_{g}\left(  \alpha\right)  Z_{q\bar{q}g}\right)  +x_{g}%
x_{q}\vec{z}_{21}K_{0}\left(  Q_{g}\left(  \alpha\right)  Z_{q\bar{q}%
g}\right)  \right)  , \nonumber%
\eqa
where%
\begin{equation}
Z_{q\bar{q}g}=\sqrt{x_{q}\frac{x_{_{\bar{q}}}+(1-\alpha)x_{q}x_{g}}%
{x_{_{\bar{q}}}+x_{q}}\vec{z}_{21}^{\,\,2}} \, .
\end{equation}
To get the previous result, we used the following parametrization
\beqa
&&\int d\vec{r}\frac{\vec{r}}{\vec{r}^{\,\,2}}e^{i(a\vec{r}^{\,\,2}+b(\vec
{r}+\vec{\rho})^{2})-i\vec{p}\cdot\vec{r}}=-2\pi\frac{i\left(  2b\vec{\rho}-\vec
{p}\right)  }{\left(  2b\vec{\rho}-\vec{p}\right)  ^{2}}e^{ib\vec{\rho
}^{\,\,2}}\left(  e^{-\frac{i\left(  2b\vec{\rho}-\vec{p}\right)  ^{2}%
}{4(a+b)}}-1\right)
\nonumber \\
&&=-\frac{\pi}{2(a+b)}\left(  2b\vec{\rho}-\vec{p}\right)  e^{ib\vec{\rho
}^{\,\,2}}\int_{0}^{1}d\alpha e^{-\alpha\frac{i\left(  2b\vec{\rho}-\vec
{p}\right)  ^{2}}{4(a+b)}}. \label{integralUnderK}%
\eqa
In the simpler case $\vec{p}_{q}=0$ the integral can be fully reduced to
\begin{equation}
\int d\vec{z}_{3}\frac{\vec{z}_{32}}{\vec{z}_{32}^{\,\,2}}K_{0}\left(
QZ_{123}\right)  =-\frac{2\pi}{x_{g}x_{q}Q}\frac{\vec{z}_{21}}{\vec{z}%
_{21}^{\,\,2}}\left(  Z_{q\bar{q}}K_{1}\left(  QZ_{q\bar{q}}\right)
-Z_{122}K_{1}\left(  QZ_{122}\right)  \right)  , \label{int_zK0_f}%
\end{equation}
where%
\begin{equation}
Z_{122}=\sqrt{x_{q}(x_{_{\bar{q}}}+x_{g})\vec{z}_{21}^{\,\,2}}.
\end{equation}
Using the integral (\ref{integralUnderK}), we also get%
\beqa
\label{int_zdK0}
&&\int d\vec{z}_{3}e^{-i\vec{p}_{g}\cdot\vec{z}_{3}}\frac{z_{32}^{j}}{\vec{z}%
_{32}^{\,\,2}}\frac{\partial}{\partial z_{3}^{l}}K_{0}(QZ_{123})=-\pi
e^{-i\vec{p}_{g}\cdot\vec{z}_{2}}\int_{0}^{1}d\alpha\, e^{\alpha\frac{ix_{q}(\vec
{z}_{21}\cdot\vec{p}_{g})}{x_{q}+x_{_{\bar{q}}}}}\left[  \left(  \delta
^{jl}+ip_{g}^{j}z_{12}^{\,\,l}\frac{x_{q}}{x_{_{\bar{q}}}+x_{q}}\right)
\right.
\nonumber \\
&&\times K_{0}\left(  Q_{g}\left(  \alpha\right)  Z_{q\bar{q}g}\right)
-\frac{x_{q}^{2}x_{g}}{x_{_{\bar{q}}}+x_{q}}z_{12}^{\,\,l}z_{12}^{j}%
\frac{Q_{g}(\alpha)}{Z_{q\bar{q}g}}K_{1}\left(  Q_{g}\left(  \alpha\right)
Z_{q\bar{q}g}\right)  (1-\alpha)\nonumber \\
&&
\left.  -\frac{x_{q}(ip_{g}^{l}z_{12}^{j}+ip_{g}^{j}z_{12}^{\,\,l})}%
{x_{q}+x_{_{\bar{q}}}}\alpha K_{0}\left(  Q_{g}\left(  \alpha\right)
Z_{q\bar{q}g}\right)  +\frac{ip_{g}^{l}ip_{g}^{j}}{x_{g}(x_{q}+x_{_{\bar{q}}%
})}\alpha\frac{Z_{q\bar{q}g}}{Q_{g}\left(  \alpha\right)  }K_{1}\left(
Q_{g}\left(  \alpha\right)  Z_{q\bar{q}g}\right)  \right]  . %
\eqa
As before, in the simpler case $\vec{p}_{g}=0$ the integral can be fully reduced to%
\beqa
&&\int d\vec{z}_{3}\frac{z_{31}^{j}}{\vec{z}_{31}^{\,\,2}}\frac{\partial
}{\partial z_{3}^{l}}K_{0}(QZ_{123})=-\frac{2\pi x_{q}}{x_{_{\bar{q}}}x_{g}%
}\left(  \delta^{jl}-2\frac{z_{12}^{l}z_{12}^{j}}{\vec{z}_{21}^{\,\,2}%
}\right)  \left(  \frac{K_{1}\left(  QZ_{q\bar{q}}\right)  }{QZ_{q\bar{q}}%
}-\frac{K_{1}\left(  QZ_{121}\right)  }{QZ_{121}}\right)
\nonumber \\
&&+2\pi\left(  \delta^{jl}\frac{K_{1}\left(  QZ_{121}\right)  }{QZ_{121}}%
-\frac{z_{12}^{l}z_{12}^{j}}{\vec{z}_{21}^{\,\,2}}K_{2}\left(  QZ_{121}%
\right)  \right)  . \label{int_zdK0_f}%
\eqa
The integrals (\ref{int_zK0}) and (\ref{int_zdK0}) are convergent and can be evaluated numerically.

\section*{Appendix B: Integrals necessary for Fourier transform}%
\label{Ap:B}

We here provide the set of Fourier transforms of modified Bessel functions we used in our calculation.
\beqa
&&\int\frac{d\vec{z}_{1}}{2\pi}\frac{d\vec{z}_{2}}{2\pi}e^{i[\vec{q}_{1}\cdot\vec
{z}_{1}+\vec{q}_{2}\cdot\vec{z}_{2}]}\frac{z_{1}^{j}}{\vec{z}_{1}^{\,\,2}}%
K_{0}(Q\sqrt{x_{g}x_{q}\vec{z}_{1}^{\,\,2}+x_{q}x_{_{\bar{q}}}\vec{z}%
_{21}^{\,\,2}+x_{_{\bar{q}}}x_{g}\vec{z}_{2}^{\,\,2}})
\nonumber \\
&&=\frac{i\left(  q_{2}{}^{j}x_{q}+q_{1}{}^{j}\left(  1-x_{\bar{q}}\right)
\right)  }{\left(  1-x_{\bar{q}}\right)  x_{\bar{q}}x_{g}x_{q}\left(
Q^{2}+\frac{\vec{q}_{2}{}^{2}}{x_{\bar{q}}(1-x_{\bar{q}}{})}\right)  \left(
Q^{2}+\frac{\vec{q}_{1}{}^{2}}{x_{q}}+\frac{\vec{q}_{2}{}^{2}}{x_{\bar{q}}%
}+\frac{\left(  \vec{q}_{1}+\vec{q}_{2}\right)  {}^{2}}{x_{g}}\right)
}\,.\label{int_dqzK0_123}%
\eqa
\beqa
&&\int\frac{d\vec{z}_{1}}{2\pi}e^{i\vec{q}_{1}\cdot\vec{z}_{1}}K_{0}(Q\sqrt
{x_{q}\left(  x_{g}+x_{\bar{q}}\right)  \vec{z}_{1}^{\,\,2}})=\frac{1}{\left(
1-x_{q}\right)  x_{q}Q^{2}+\vec{q}_{1}{}^{2}} \, .\label{int_dqzK0_122}%
\eqa
\beqa
\label{int_dqzz1z2K1_123}
&&\hspace{-.3cm}\int\frac{d\vec{z}_{1}}{2\pi}\frac{d\vec{z}_{2}}{2\pi}e^{i[\vec{q}_{1}\cdot\vec
{z}_{1}+\vec{q}_{2}\cdot\vec{z}_{2}]}\frac{z_{1}^{j}z_{2}^{\beta}}{\vec{z}_{2}%
{}^{2}}\frac{K_{1}(Q\sqrt{x_{g}x_{q}\vec{z}_{1}^{\,\,2}+x_{q}x_{_{\bar{q}}%
}\vec{z}_{21}^{\,\,2}+x_{_{\bar{q}}}x_{g}\vec{z}_{2}^{\,\,2}})}{\sqrt
{x_{g}x_{q}\vec{z}_{1}^{\,\,2}+x_{q}x_{_{\bar{q}}}\vec{z}_{21}^{\,\,2}%
+x_{_{\bar{q}}}x_{g}\vec{z}_{2}^{\,\,2}}}%
 \\
&&\hspace{-.3cm}=\frac{x_{\bar{q}}\left(  \delta^{j\beta}-\frac{2\left(  q_{2}{}^{j}\left(
x_{\bar{q}}+x_{g}\right)  +q_{1}{}^{j}x_{\bar{q}}\right)  \left(  q_{2}%
{}^{\beta}\left(  x_{\bar{q}}+x_{g}\right)  +q_{1}{}^{\beta}x_{\bar{q}%
}\right)  }{(\vec{q}_{2}{}\left(  x_{\bar{q}}+x_{g}\right)  +\vec{q}_{1}%
{}x_{\bar{q}})^{2}}\right)  }{2Qx_{q}(\vec{q}_{2}{}\left(  x_{\bar{q}}%
+x_{g}\right)  +\vec{q}_{1}{}x_{\bar{q}})^{2}}\ln\left(  \frac{\frac{q_{2}%
{}^{2}}{x_{\bar{q}}}+\frac{\left(  q_{1}+q_{2}\right)  {}^{2}}{x_{g}}%
+\frac{q_{1}{}^{2}}{x_{q}}+Q^{2}}{\frac{q_{1}{}^{2}}{x_{q}\left(
1-x_{q}\right)  }+Q^{2}}\right)\nonumber
\\
&&\hspace{-.3cm}+\frac{q_{2}{}^{\beta}\left(  x_{\bar{q}}+x_{g}\right)  +q_{1}{}^{\beta
}x_{\bar{q}}}{Qx_{q}^{2}(\vec{q}_{2}{}\left(  x_{\bar{q}}+x_{g}\right)
+\vec{q}_{1}{}x_{\bar{q}})^{2}}\!\!\left( \! \frac{q_{1}{}^{j}\left(  x_{g}%
+x_{q}\right)  +q_{2}{}^{j}x_{q}}{x_{g}\!\left( \! \frac{q_{2}{}^{2}}{x_{\bar{q}}%
}+\frac{\left(  q_{1}+q_{2}\right)  {}^{2}}{x_{g}}+\frac{q_{1}{}^{2}}{x_{q}%
}+Q^{2}\!\right)  }\!-\!\frac{q_{1}{}^{j}}{\left(  x_{\bar{q}}+x_{g}\right) \! \left(
\frac{q_{1}{}^{2}}{x_{q}\left(  x_{\bar{q}}+x_{g}\right)  }+Q^{2}\right)
}\!\right)  \!. \nonumber%
\eqa
\beqa
&&\int\frac{d\vec{z}_{1}}{2\pi}\frac{d\vec{z}_{2}}{2\pi}e^{i[\vec{q}_{1}\cdot\vec
{z}_{1}+\vec{q}_{2}\cdot\vec{z}_{2}]}\frac{z_{2}^{j}z_{2}^{\beta}}{\vec{z}_{2}%
{}^{2}}\frac{K_{1}(Q\sqrt{x_{g}x_{q}\vec{z}_{1}^{\,\,2}+x_{q}x_{_{\bar{q}}%
}\vec{z}_{21}^{\,\,2}+x_{_{\bar{q}}}x_{g}\vec{z}_{2}^{\,\,2}})}{\sqrt
{x_{g}x_{q}\vec{z}_{1}^{\,\,2}+x_{q}x_{_{\bar{q}}}\vec{z}_{21}^{\,\,2}%
+x_{_{\bar{q}}}x_{g}\vec{z}_{2}^{\,\,2}}}%
\nonumber \\
&&=\frac{\left(  x_{\bar{q}}+x_{g}\right)  \left(  \delta^{j\beta}%
-\frac{2\left(  q_{2}{}^{j}\left(  x_{\bar{q}}+x_{g}\right)  +q_{1}{}%
^{j}x_{\bar{q}}\right)  \left(  q_{2}{}^{\beta}\left(  x_{\bar{q}}%
+x_{g}\right)  +q_{1}{}^{\beta}x_{\bar{q}}\right)  }{(\vec{q}_{2}{}\left(
x_{\bar{q}}+x_{g}\right)  +\vec{q}_{1}{}x_{\bar{q}})^{2}}\right)  }%
{2Qx_{q}(\vec{q}_{2}{}\left(  x_{\bar{q}}+x_{g}\right)  +\vec{q}_{1}{}%
x_{\bar{q}})^{2}}\ln\left(  \frac{\frac{q_{2}{}^{2}}{x_{\bar{q}}}%
+\frac{\left(  q_{1}+q_{2}\right)  {}^{2}}{x_{g}}+\frac{q_{1}{}^{2}}{x_{q}%
}+Q^{2}}{\frac{q_{1}{}^{2}}{x_{q}\left(  1-x_{q}\right)  }+Q^{2}}\right)
\nonumber \\
&&+\frac{\left(  q_{2}{}^{j}\left(  x_{\bar{q}}+x_{g}\right)  +q_{1}{}%
^{j}x_{\bar{q}}\right)  \left(  q_{2}{}^{\beta}\left(  x_{\bar{q}}%
+x_{g}\right)  +q_{1}{}^{\beta}x_{\bar{q}}\right)  }{Qx_{g}x_{q}x_{\bar{q}%
}(\vec{q}_{2}{}\left(  x_{\bar{q}}+x_{g}\right)  +\vec{q}_{1}{}x_{\bar{q}%
})^{2}\left(  \frac{q_{2}{}^{2}}{x_{\bar{q}}}+\frac{\left(  q_{1}%
+q_{2}\right)  {}^{2}}{x_{g}}+\frac{q_{1}{}^{2}}{x_{q}}+Q^{2}\right)
}\,.\label{int_dqzz2z2K1_123}%
\eqa
\begin{equation}
\int\frac{d\vec{z}_{1}}{2\pi}e^{i\vec{q}_{1}\vec{z}_{1}}z_{1}^{j}\frac
{K_{1}(Q\sqrt{x_{q}\left(  x_{g}+x_{\bar{q}}\right)  \vec{z}_{1}^{\,\,2}}%
)}{\sqrt{x_{q}\left(  x_{g}+x_{\bar{q}}\right)  \vec{z}_{1}^{\,\,2}}}%
=\frac{iq_{1}^{j}}{\left(  1-x_{q}\right)  x_{q}Q\left(  \left(
1-x_{q}\right)  x_{q}Q^{2}+\vec{q}_{1}{}^{2}\right)  } \,.\label{int_dqzz1K1_122}%
\end{equation}

\providecommand{\href}[2]{#2}\begingroup\raggedright\endgroup

\end{document}